\documentclass[a4paper,fleqn,usenatbib]{mnras}

\usepackage{newtxtext,newtxmath}
\usepackage[T1]{fontenc}
\usepackage{ae,aecompl}

\usepackage{graphicx}	% Including figure files
\usepackage{amsmath}	% Advanced maths commands
\title[156 Reliable Nova Periods]{Comprehensive Listing of 156 Reliable Orbital Periods for Novae, Including 49 New Periods}

\author[B. E. Schaefer]{
Bradley E. Schaefer$^{1}$\thanks{E-mail: schaefer@lsu.edu},
\\
$^{1}$Department of Physics and Astronomy, Louisiana State University, Baton Rouge, Louisiana, 70820, USA\\
}

\pubyear{2021}

\begin{document}
\label{firstpage}
\pagerange{\pageref{firstpage}--\pageref{lastpage}}
\maketitle

\begin{abstract}

I report on a large-scale search for the orbital periods ($P$) of most known nova systems, by looking for significant, coherent, and stable optical photometric modulation in two or more independent light curves taken mostly from the large surveys of $TESS$, $Kepler$, AAVSO, SMARTS, OGLE, ASAS, and ZTF.  I have discovered 31 new orbital periods.  Further, I have measured new periods for 18 novae with evolved companions, to 30 per cent accuracy, as based on their spectral energy distribution.  Also, I have confirmed, improved, and rejected prior claims for $P$ in 46 novae.  (As part of this effort, I recognize that 5 novae display 1--3 coherent, significant, and transient periodicities 0.12--4.1 days, with these being mysterious as not being the orbital, spin, or superhump periods.)  In all, I have compiled a comprehensive list of 156 {\it reliable} $P$ values for novae.  The histogram of nova periods shows a minimum $P$ at 0.059 hours (85 minutes), and a Period Gap from 0.071--0.111 days (1.70--2.66 hours).  The upper edge of the Period Gap is significantly different between novae (0.111 days), nova-like systems (0.131 days), and dwarf novae (0.141 days).  A further issue from the histogram is that 31 per cent of nova systems have evolved companions, for which there has been no models or understanding for their current state or evolution.  For the novae with red giant companions, 15-out-of-20 are in the bulge population, despite novae with main-sequence and subgiant companions having bulge fractions near 0.11--0.32.
 
\end{abstract}


%%%%%%%%%%%%%%%%%%%%%%%%%%%%%%%%%%%%%%%%%%%%%%%%%%

%%%%%%%%%%%%%%%%% BODY OF PAPER %%%%%%%%%%%%%%%%%%

\section{Introduction}

Galactic novae appear as stars that suddenly brighten by over seven magnitudes in hours-to-weeks from a faint long-standing quiescence to a very-luminous peak that lasts for days-to-months before fading slowly over months-to-years back to near the pre-eruption level (Payne-Gaposchkin 1964).  Novae are the classical novae (CNe) and recurrent novae (RNe), which are cataclysmic variables (CVs) composed of a relatively-ordinary companion star circling a white dwarf, with mass falling off the companion on to the white dwarf, where it accumulates until the pressure is great enough to trigger a runaway thermonuclear explosion.  Most nova binaries have orbital periods ($P$) from 3--8 hours, yet with a total range from 0.059 to 748 days.

The orbital period is the single most important parameter of any nova, as it determines the evolutionary state, the nature of the companion, and is required for much of the physical modeling.  So, for many decades, our community has been using vast amounts of telescope time to discover $P$ for as many novae as possible.  This large program started with M. Walker and R. Kraft discovering the eclipses of DQ Her and other novae (Walker 1954; Kraft 1964).  Always excellent programs at Dartmouth (centred on J. Thorstensen), at San Diego State University (centred on A. W. Shafter), and at the University of Capetown (centred on B. Warner and P. A. Woudt) have each produced many reliable periods.  In one impressive paper, Mr\'{o}z et al. (2015) used their huge OGLE data base to discover 19 nova periods from targets near the galactic centre.  A large group at several Chilean institutions (centred around C. Tappert) has since 2012 been putting out an impressive series of eight papers in the {\it Monthly Notices} where they track down and prove identifications of the quiescent counterparts, with this work netting 18 $P$ discoveries.

This huge period-search enterprise is important both for understanding and modeling the physics of individual systems, but also of broad importance for seeing the evolution of CVs.  The characteristic analysis is to use a compilation of $P$ values to find the minimum orbital period ($P_{\rm min}$) and to define the famous Period Gap (ranging from $P_{\rm gap,-}$ to $P_{\rm gap,+}$) with few systems.  Among compilations of nova $P$ measures, the old standard catalog was the {\it General Catalog of Variable Stars} (GCVS), but this was always sparse and is now long out of date (Samus et al. 2017).  The replacement for the GCVS is the {\it International Variable Star Index} (VSX) run by the {\it American Association of Variable Star Observers} (AAVSO), providing comprehensive information of many types for all variable stars, with this now being the best and official source for CV information.  The Catalog and Atlas of Cataclysmic Variables (CV-Cat) is an ever-useful compilation of all novae up until 2006, their orbital periods, their basic properties, and great finder charts{\footnote{https://archive.stsci.edu/prepds/cvcat/index.html}} (Downes et al. 2001).  H. Ritter and U. Kolb have compiled a list specializing in orbital periods of all CVs (Ritter \& Kolb 2003), with this being updated to 2015{\footnote{https://wwwmpa.mpa-garching.mpg.de/RKcat/}}.

Up until now, the period-search enterprise has required very long hours over many nights and years at telescopes.  But an exciting non-traditional period-search method has just opened up.  The idea is to use existing public on-line data bases that contain extensive and accurate photometry of most nova systems in the sky.  The task is essentially to take discrete Fourier transforms (DFTs) of these wonderful light curves, then recognize coherent and stable periodic photometric modulations.  Most nova binaries will show a photometric modulation exactly at the orbital period.  Any periodicity in the period range 0.04--10 days that is coherent, stable, and significant must be tied to a very good clocking mechanism in the binary, and that can only be the orbital period.

The primary public data bases are for the {\it Kepler} satellite (their K2 mission in particular), the {\it Transiting Exoplanet Survey Satellite} ($TESS$) now on-going, and the {\it Zwicky Transient Facility} (ZTF) that is also on-going.  Further useful sources are the photometry in the International Database of the {\it American Association of Variable Star Observers} (AAVSO), the impressive photometric and spectroscopic  Stony Brook/SMARTS Atlas of (mostly) Southern Novae (SMARTS) run by F. Walter (Walter et al. 2012), the {\it All Sky Automated Survey} (ASAS), and the {\it Optical Gravitational Lensing Experiment} (OGLE) .  Roughly, these surveys cover nearly all novae in the sky down to $\sim$19 mag with extensive well-sampled light curves.

This new non-traditional method has only become possible in the last few years, yet our nova community has so far made only scant use.  With the realization of the power of this method, I have systematically searched through all known galactic novae, trying to find new nova $P$.  This task could not proceed far for the substantial number of old novae that are too faint in quiescence to have useful coverage from any of the new surveys.  For my period searches, I require a confident proof of the orbital period, with the primary method being to measure the same periodicity in two or more independent light curves, requiring each to have the signal as coherent, stable, and significant.  The result is that I have discovered 31 new orbital periods for novae.  The first part of this paper consists of reporting on my 31 new nova periods.  Further, I identify 24 novae with evolved companion stars as based on clear blackbody shapes in their spectral energy distributions (SEDs), and I then derive 18 new $P$ values as based on the calculated blackbody radii of the companion stars plus Kepler's Law.  These orbital periods are reliable, albeit with typically 30 per cent accuracy, adequate for many purposes.  My 49 new periods now constitutes over 31 per cent of all known reliable nova $P$ values.

The second part of this paper is to collect all known reliable nova $P$.  This entails me going deep into the literature and into the modern survey databases for over 200 novae, collecting the best periods.    I used the modern public survey data bases to confirm, improve, and reject prior claimed periods for 46 novae.  My result is that I have collected a list of 156 nova $P$ values that are reliable.  This nearly doubles the best prior compilation from Fuentes-Morales et al. (2021), which reports 92 orbital periods yet with 13 periods now-known to be greatly wrong.  With my much-larger and purer set of periods, I can now resolve the Period Gap with good resolution.  Further, a perhaps-surprising realization is that substantial fractions of the nova population are below the Period Gap (5 novae for 3.2 per cent), are inside the Period Gap (5 novae for 3.2 per cent), are with subgiant companions (28 novae for 18 per cent), and are with red giant companions (20 novae for 13 per cent).

\section{New Orbital Periods}

\subsection{Light Curve Data}

$TESS$ is a satellite mission that covers much of the sky down to roughly 20th mag with one-or-more sectors of data lasting up to 27 days with continuous photometry with cadences from 20--1800 seconds (Ricker et al. 2015).  Each $TESS$ sector consists of data from two orbits, separated by a one-day gap.  The $TESS$ data are processed, stored, and distributed to the public by the Mikulski Archive for Space Telescopes (MAST){\footnote{https://mast.stsci.edu/portal/Mashup/Clients/Mast/Portal.html}}.  My primary tool for extracting light curves has been the {\sc Lightkurve} package (Lightkurve Collaboration, 2018).

The {\it Kepler} spacecraft made $\sim$67 day continuous stares as part of its K2 mission, with 20--1800 second integration times, covering many novae around the crowded galactic centre (Howell et al. 2014).  These data are publicly available at the MAST website.

The AAVSO International Database{\footnote{https://www.aavso.org/aavso-international-database-aid}} contains over a million magnitude estimates of novae alone, with most in the last twenty or so years being from CCD photometry with professional quality. Therefore my first look at any nova light curve is always to check out the AAVSOs Light Curve Generator (LCG){\footnote{https://www.aavso.org/LCGv2/}}.  All data are publicly available for downloading at an AAVSO website{\footnote{https://www.aavso.org/data-download}}.

The Stony Brook / SMARTS Spectral Atlas of Southern Novae (SMARTS) has extensive long-term spectroscopy (with both high- and low-resolution), {\it BVRIJHK} photometry, plus finder charts for 114 southern novae with eruptions from 2003 to the present, both in quiescence and in eruption, from Cerro Tololo (Walter et al. 2012).  The SMARTS data are publicly available at a Stony Brook web site{\footnote{http://www.astro.sunysb.edu/fwalter/SMARTS/NovaAtlas/}}.

OGLE covers much of the galactic centre region down to 19.5 mag from 2001-2015, with their nova photometry presented in Mr\'{o}z et al. (2015).  The OGLE nova light curve are publicly available\footnote{http://ogle.astrouw.edu.pl/ogle/ogle4/NOVAE/}.

ASAS covers the entire sky down to roughly 14th mag, typically with hundreds of magnitudes from 2000 to 2009 (Pojmanski 1997).  Their light curves are publicly available for download{\footnote{http://www.astrouw.edu.pl/asas/?page=aasc}}.

The ZTF survey{\footnote{https://irsa.ipac.caltech.edu/cgi-bin/Gator/nph-scan?projshort=ZTF} covers the entire sky north of $-$29$\degr$ declination to a depth of 20.5 mag in two bands ($zg$ and $zr$), with most stars covered by several hundreds of magnitudes since 2018, with the survey now on-going (Bellm et al. 2019).

The raw light curves usually need some sort of corrections.  For example, the reported times must all be converted to heliocentric Julian dates (HJDs).  The $TESS$ and K2 missions report times for the reference frame of Barycentric Julian Date (BJD), which for purposes in this paper is negligibly different from HJD.  Some light curves are detrended (pre-whitened) to remove slow variations, such as arise in the fading tail of the nova eruption and as arise from imperfectly corrected background light in the $TESS$ light curves.  Such detrending takes out any signal from long periods (but such would never be reliable in any case), while the not-slow variations (typically faster than one-day) will come out in the DFT with little alteration.  Another common correction is to normalize different data sets to the same average magnitude to create a single joint light curve for DFT purposes.  A typical case is to shift the ZTF $zg$ and $zr$ light curves to a common mean, with substantial improvement in the joint DFT.  Another common need is to recognize and toss out the various outliers that always appear in light curves, whether ground-based or space-based.  Examples of this are from outliers associated with thruster firings on the $Kepler$ spacecraft, and the first-and-last few hours of a segment of $TESS$ light curves.

\subsection{Discovery of New Orbital Periods}

The primary analysis tool is the usual discrete Fourier transform (DFT).  This has the big advantage over other periodogram methods because amplitudes, error bars, significances, and alias structures are well-known and standard.  My basic search is for periods from 0.04 days (set to just under the shortest known orbital period for any nova) to 10 days (set by the length of the TESS light curves).  Sample DFTs for four novae are presented in Fig. 1.

\begin{figure}
	\includegraphics[width=1.01\columnwidth]{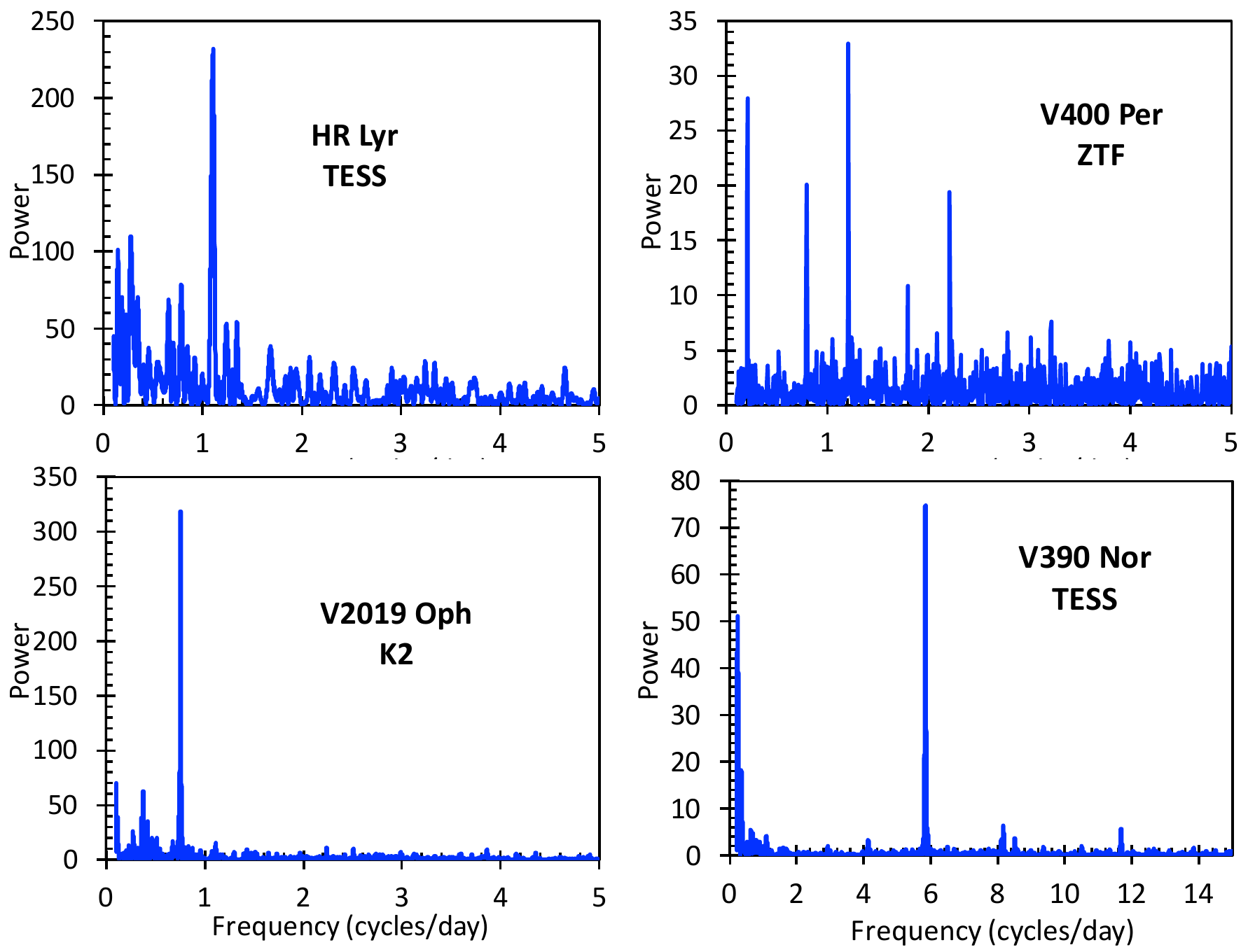}
    \caption{DFTs for four novae.  The frequency is in units of cycles/day ($P$ is the inverse of frequency), while the Fourier power is in units of the average DFT power.  This sampling shows typical DFTs from $TESS$ (HR Lyr, upper left panel), ZTF (V400 Per, upper right panel), and K2 (V2109 Oph, lower left panel).  All of the DFTs are just variations on these, as described in the text for individual stars.  For $TESS$, we see the typical low frequency noise caused by imperfect detrending of systematic problems with standard programs from MAST.  For ZTF, we see the typical problems with daily aliases caused by the observations all being taken from one longitude.  The DFT in the lower right panel (for V390 Nor) is one of the poorest cases amongst the DFTs for the new periods in Table 1,  with the smallest apparent amplitude of modulation from only one $TESS$ sector, and where the usual low-frequency noise from poor detrending is prominent.}
\end{figure}

For my new nova periods, I require that each is detected significantly in {\it two} or more {\it independent} data sets.  This is a significant burden.  However, this requirement is my primary means to prove the reliability of the orbital period.  That is, almost all artefacts and randomness cannot give identical periodicities (in terms of periods, epochs, and amplitudes) in independent light curves.  The most common case of miscues in the literature is when a relatively scanty light curve is dominated by a a small number of shot-noise peaks (e.g., flickering) whereupon the DFTs will always display peaks that are nominally significant, as based on calculations where the light curve noise is uncorrelated and Gaussian.  Another common miscue is to look over a relatively short time span and interpret the usual semi-regular variations as being the orbital period, whereas over every different interval shows greatly-different apparent periodicities.  These problems are all solved when the same periodic signal is detected significantly in independent data sets from different times.  

%The significance of a periodic modulation can be measured by several means:  First, the probability that a DFT peak exceeds a power of $\mathcal{P}$ is e$^{\mathcal{P}/\langle \mathcal{P} \rangle}$, where $\langle \mathcal{P} \rangle$ is the local average noise power, for assumed uncorrelated noise.  For a period search over a  range of frequencies ($\Delta$F, say with 0.1 to 25 cycles/day as in my search) and for peaks with a FWHM width of $\sigma_F$, the probability must be corrected by a factor of $\Delta$F/$\sigma_F$ to account for the number of independent trial periods searched.  A second standard method to test the significance of a periodicity is to perform a chi-square fit, vary the one amplitude parameter to zero, where the difference, i.e., $\chi^2_{\rm Amp=0}$-$\chi^2_{\rm min}$, will be the square of the Gaussian significance.  A third method is to look at the DFT and see whether the peak is isolated and far above all other noise peaks.  This does not count related alias peaks, known peaks from the window function, and low-frequency noise arising from irrelevant trends.  This informal method is fast, easy, and reliable for recognizing highly significant periods.

A substantial problem for $\it TESS$ light curves is that the standard programs and analysis from {\it MAST} will occasionally remove true periodicities and will occasionally insert false periodicities.  The various {\sc Lightkurve} and {\it MAST} data products have variability components removed as optimized for the visibility of transits, with this sometimes deleting periodic signals, even of large amplitude.  A partial defense for this problem is to extract the light curve with multiple methods, where hopefully they do not all remove the sought signal.  The usual mode for the creation of false periodicities is by slightly imperfect background subtraction, where structure in the background then appears in the official light curves, and this structure can then appear as apparently-significant DFT peaks coming from the the coincidences of time intervals between peaks and dips.  I have not found any notice or documentation of these artefacts, and the effects change over time as the programs are changed.  One good defense is to require the same periodicity to appear in multiple light curves, as the artefacts are unlikely to repeat.  Another good defense is to construct otherwise identical light curves for nearby stars and for nearby sky regions, and to see what DFT peaks arise.

My requirement that each new periodicity must appear significantly in two-or-more independent data sets provides a guarantee of reliability for the existence of the periodicity.  My further requirement is that the periodicity be coherent over a large number of cycles, which ties the modulation to an accurate clock mechanism, which for periods from 0.04--10 days can only be the orbital period.

Once a period is identified, I fitted a periodic light curve template to the original light curve.  In most cases a simple sinewave was consistent with the best folded light curve, with the only free parameters in the fit being the period, the epoch of minimum light (i.e., when the star is at its faintest), and the amplitude.  For some stars, the best folded light curves show prominent secondary minima and unequal magnitudes at the orbital elongations, and for these I always used a designed template shape that it close to what is seen from the nova.

Fig. 2 shows four examples of folded light curves.  The upper right panel (for V356 Aql) shows a typical case where the light curve is consistent with a sinewave.  For 28 of the 31 novae with new periods, the folded light curve only shows a sinusoid, and these all look alike, with the underlying information contained in the best fitting parameters (see Table 1).  The amplitude of the scatter arises from ordinary flickering, measurement and Poisson errors, the ubiquitous ups-and-downs on long time scales, and imperfect detrending from systematic problems.  So the scatter in these folded light curves can often be quite large, with essentially no relation to the real modulation of the star.  Fortunately, the light curves usually have thousands or tens-of-thousands of points, so the scatter in the light curve is greatly reduced when looking at the folded light curve with averaging over phase-bins.  The remaining three panels display the folded light curves for the three novae that show significant structure more complicated than a sinewave (i.e., GI Mon, V697 Sco, and V1186 Sco).  These three light curves show secondary eclipses, and are typical for novae.

\begin{figure}
	\includegraphics[width=1.01\columnwidth]{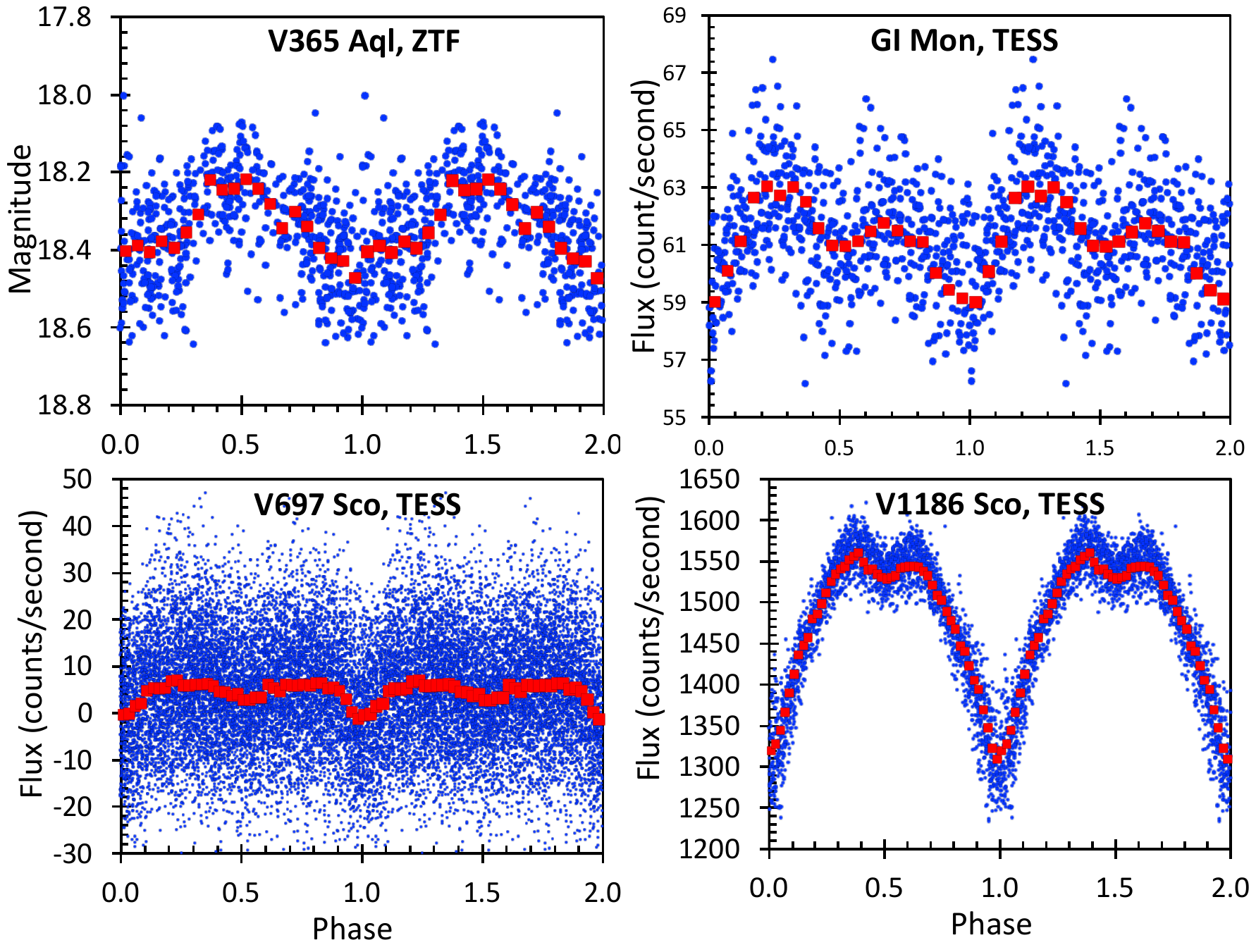}
    \caption{Folded light curves for four novae.  The individual magnitudes or fluxes in the light curve (small blue circles) are folded on the orbital phase (with double plotting from 1--2 in phase).  The phase-bin averaged values are plotted as red squares, for which the error bars are smaller than the plot symbol.  The upper left panel (for V365 Aql) is typical of the 28 other cases in Table 1 for which the folded light curve is not significantly different from a sinewave.  The three novae (GI Mon, V697 Sco, and V1186 Sco) are the only novae from Table 1 for which the shape is different from a sinusoid.}
\end{figure}

The fitted amplitudes reported for $TESS$ and K2 have the problem that the photometry apertures always contain many stars in addition to the nova.  This means that the measured fluxes often have a large constant added to the flux from the nova.  With this, the calculated light curve amplitudes, as reported in magnitudes, will always be smaller than for the nova alone.  I have no accurate or practical means to correct for this effect.

\subsection{Details For Each New Period}

I have discovered 31 new reliable orbital periods for novae.  Table 1 lists the details of the new measures.  The first three columns give the nova name (in GCVS order), the year of peak, and the light curve class (from Strope, Schaefer, \& Henden 2010).  The next three columns give the new $P$ in days, the epoch of minimum light in HJD, and the full amplitude (i.e., peak-to-peak) of modulation in units of magnitudes.  An asterisk after the amplitude means that the value should be greatly larger due to the inclusion of substantial extra light in the photometry aperture from ordinary background stars.  The last two columns give the sources, year range of data, and the number of light curve points, for each of the two data sets in which the periodicity is significant.

\begin{table*}
	\centering
	\caption{31 Novae With Newly Discovered Periods}
	\begin{tabular}{llllllll} 
		\hline
		Nova & Year &  LC    & $P$ (days)  &  Epoch Minimum (HJD)  &   Ampl.   &   Dataset 1 (year, \#)  &  Dataset 2 (year, \#)   \\
		\hline
V356 Aql	&	1936	&	J(140)	&	0.4265059	$\pm$	0.0000067	&	2458911.0749	$\pm$	0.0043	&	0.18	&	ZTF (2019, 299)	&	ZTF (2020-21, 287)	\\
V1370 Aql	&	1982	&	D(29)	&	1.95810	$\pm$	0.00017	&	2458816.748	$\pm$	0.036	&	0.17	&	ZTF (2018-19, 470)	&	ZTF (2020-21, 336)	\\
V1405 Cas	&	2021	&	J(175)	&	0.1883907	$\pm$	0.0000048	&	2458859.0688	$\pm$	0.0021	&	0.003*	&	$TESS$ (2019, 1831)	&	$TESS$ (2020, 1167)	\\
V1369 Cen	&	2013	&	D(65)	&	0.156556	$\pm$	0.000024	&	2458611.0079	$\pm$	0.0012	&	0.009*	&	$TESS$ (2019, 585)	&	$TESS$ (2019, 597)	\\
FM Cir	&	2018	&	J(85)	&	0.1497672	$\pm$	0.0000309	&	2459053.0348	$\pm$	0.0018	&	0.0025	&	$TESS$ (2019, 2351)	&	$TESS$ (2021, 3411)	\\
V339 Del	&	2013	&	PP(21)	&	0.162941	$\pm$	0.000060	&	2458696.1286	$\pm$	0.0028	&	0.020	&	$TESS$ (2019, 14167)	&	$TESS$ (2021, 107805)	\\
KT Eri	&	2009	&	PP(14)	&	2.61595	$\pm$	0.000060	&	2455491.323	$\pm$	0.053	&	0.35	&	$TESS$ (2020, 102757)	&	SMARTS (2010-11, 109)	\\
V407 Lup	&	2016	&	S(8)	&	3.62	$\pm$	0.05	&	2458641.8788	$\pm$	0.0515	&	0.003	&	$TESS$ (2019, 1111)	&	$TESS$ (2021, 3464)	\\
HR Lyr	&	1919	&	S(97)	&	0.905778	$\pm$	0.000016	&	2459023.5540	$\pm$	0.0064	&	0.052*	&	$TESS$ (2019-20, 1259)	&	$TESS$ (2021, 19610)	\\
GI Mon	&	1918	&	S(23)	&	0.4470645	$\pm$	0.0000008	&	2459197.1630	$\pm$	0.0009	&	0.078*	&	$TESS$ (2019, 1072)	&	$TESS$ (2021, 16722)	\\
QY Mus	&	2008	&	S(95)	&	0.901135	$\pm$	0.000026	&	2459190.6139	$\pm$	0.0087	&	0.0013	&	$TESS$ (2019, 897)	&	$TESS$ (2021, 3282)	\\
V357 Mus	&	2018	&	D(32)	&	0.155163	$\pm$	0.000033	&	2458598.146	$\pm$	0.007	&	0.001*	&	$TESS$ (2019, 2268)	&	$TESS$ (2021, 5863)	\\
V390 Nor	&	2007	&	J(118)	&	0.171326	$\pm$	0.000042	&	2459375.1629	$\pm$	0.0016	&	0.004*	&	$TESS$ (2021, 1540)	&	$TESS$ (2021, 1451)	\\
V2109 Oph	&	1969	&	...	&	1.32379	$\pm$	0.00048	&	2457693.0927	$\pm$	0.0077	&	0.199	&	K2 (2016, 1048)	&	K2 (2016, 2027)	\\
V2487 Oph	&	RN	&	P(8)	&	1.24	$\pm$	0.02	&	2457537.394	$\pm$	0.004	&	0.04	&	CT+McD (2002-21, 1809)	&	K2 (2016, 97400)	\\
V2574 Oph	&	2004	&	S(41)	&	0.1350862	$\pm$	0.0000046	&	2457694.0295	$\pm$	0.0007	&	0.061*	&	K2 (2016, 32913)	&	K2 (2016, 65742)	\\
V392 Per	&	2018	&	P(11)	&	3.21997	$\pm$	0.00039	&	2459135.4580	$\pm$	0.0095	&	0.122	&	AAVSO (2019-21, 28725)	&	$TESS$ (2019, 1124)	\\
V400 Per	&	1974	&	S(43)	&	0.826387	$\pm$	0.000043	&	2458814.729	$\pm$	0.016	&	0.16	&	ZTF (2018-9, 412)	&	ZTF (2020-21, 247)	\\
HS Pup	&	1963	&	S(65)	&	0.178641	$\pm$	0.000044	&	2458978.1706	$\pm$	0.0012	&	0.013*	&	$TESS$ (2019, 1877)	&	$TESS$ (2021, 3299)	\\
V598 Pup	&	2007	&	...	&	0.162874	$\pm$	0.000036	&	2459025.0497	$\pm$	0.0011	&	0.041	&	$TESS$ (2018-19, 2060)	&	$TESS$ (2020-21, 34347)	\\
YZ Ret	&	2020	&	P(22)	&	0.1324539	$\pm$	0.0000098	&	2458408.0161	$\pm$	0.0013	&	0.031	&	$TESS$ (2018, 1201)	&	$TESS$ (2018, 853)	\\
GR Sgr	&	1924	&	...	&	29.4956	$\pm$	0.0040	&	2457517.32	$\pm$	0.28	&	0.34	&	ZTF (2018-21, 392)	&	OGLE (2001-21, 210)	\\
V5558 Sgr	&	2007	&	J(157)	&	0.185808	$\pm$	0.000008	&	2457539.0607	$\pm$	0.0008	&	0.003*	&	K2 (2016, 1015)	&	K2 (2016, 1874)	\\
V697 Sco	&	1941	&	...	&	1.26716	$\pm$	0.00050	&	2459375.5752	$\pm$	0.0009	&	$>$1*	&	$TESS$ (2021, 9413)	&	$TESS$ (2021, 8871)	\\
V719 Sco	&	1950	&	D(24)	&	0.43639	$\pm$	0.00039	&	2459375.2801	$\pm$	0.0054	&	0.002*	&	$TESS$ (2021, 1540)	&	$TESS$ (2021, 1451)	\\
V1186 Sco	&	2004	&	J(62)	&	0.202968	$\pm$	0.000002	&	2457693.1246	$\pm$	0.0002	&	0.182	&	K2 (2016, 2651)	&	$TESS$ (2019-21, 3772)	\\
V373 Sct	&	1975	&	J(79)	&	0.819099	$\pm$	0.000016	&	2458611.4021	$\pm$	0.0051	&	0.29	&	ZTF (2018-19, 611)	&	ZTF (2020-1, 152)	\\
XX Tau	&	1927	&	D(42)	&	0.1293567	$\pm$	0.0000011	&	2458890.0624	$\pm$	0.0025	&	0.214	&	ZTF (2018-21, 337)	&	$TESS$ (2020, 3425)	\\
V549 Vel	&	2017	&	J(118)	&	0.4031692	$\pm$	0.0000009	&	2459087.1341	$\pm$	0.0009	&	0.012*	&	$TESS$ (2019, 2035)	&	$TESS$ (2021, 5671)	\\
NQ Vul	&	1976	&	D(50)	&	0.1462568	$\pm$	0.0000006	&	2459351.0874	$\pm$	0.0008	&	0.007*	&	$TESS$ (2019, 638)	&	$TESS$ (2021, 5800)	\\
PW Vul	&	1984	&	J(116)	&	0.1285753	$\pm$	0.0000007	&	2459490.7978	$\pm$	0.0015	&	0.112	&	ZTF (2018-21, 828)	&	$TESS$ (2021, 3657)	\\
		\hline
	\end{tabular}

\end{table*}

{\bf V356 Aql}~~The ZTF light curve shows a simple sinewave that is highly significant and coherent in each year from 2019 to 2021 and in the two bands ($zg$ and $zr$).  There is no periodic signal in 2018, with the star being in a `low-state' (one magnitude fainter than in 2019--2021).  The data from the single site in California are restricted to 0.37 day in the sidereal time, and this makes for significant daily alias peaks in the DFT.  In particular, the DFT shows the four highest peaks at periods of 0.426, 0.745, 1.508, and 2.939 days, with these all being ordinary daily aliases.  Therefore, the single intrinsic periodicity (significant, stable, and coherent) must be orbital, and it must be one of those four periods.  The four peaks have similar peak powers in the DFT, as appropriate for noisy light curves made over such a small range of sidereal times.  In this case (with long flickers that are comparable in amplitude to the periodic modulation, and with relatively few nights of coverage), there is substantial random fluctuations in the power for each peak, and for other indicators that might have been able to distinguish amongst the four aliases.  The three longest periods can be confidently recognized as being aliases, as based on the system's absolute magnitude of $M_V$=$+$8.9 mag.  (The $Gaia$ parallax is 1.72$\pm$0.37 milli-arcseconds, the $V$ magnitude in quiescence is 18.3, and the $E(B-V)$ is near 0.2 mag.)  If the system has an evolved companion star (as required by the large stellar size forced by the presumed long orbital period), then the companion's luminosity would be much too bright to allow for an $M_V$ of $+$8.9 mag.  Indeed, any system with a 0.745 day period, or longer, to have the subgiant companion to be much more luminous than $+$8.9 mag, and even a 0.426 day period is pushing it.  A chi-square fitting to a sinewave of the 2019--2021 ZTF light curve gives a period is 0.4265059$\pm$0.0000067 days, and full amplitude of 0.18 mag.  The folded light curve is shown in Fig. 2.

{\bf V1370 Aql}~~The ZTF light curve has a DFT with a highly significant peak at a period of 1.958 days, far above the noise level, that appears in all combinations of $zg$ and $zr$ bands and the various years from 2018 to 2021.  Daily alias peaks (with the highest at 2.032 days) are present, but they are all greatly lower in peak power.  This periodicity is coherent, stable, and significant.  A chi-square fit of a sinewave to the ZTF light curve gives a period of 1.95810$\pm$0.00017 with a full amplitude of 0.17 mag.

{\bf V1405 Cas}~~The pre-nova counterpart was recognized by Z. Henzl {\it before} its eruption in 2021, being reported as a EW class eclipsing star with a period of 0.376938 days and a name of CzeV 3217 Cas.  But an EW eclipsing binary does not have any white dwarf to make a nova event.  Without knowing that the system is a nova, it was easy and reasonable to class the roughly-double-sinewave light curve as an EW star with twice the orbital period of the underlying nova.  Now, with the excellent $TESS$ pre-eruption fluxes, the folded light curve is a simple sinewave with flickering at a period 0.1883907 days.  In particular, this modulation is seen with the same epoch, period, and amplitude for $TESS$ sector 17 (starting 2019 October 8), sector 18 (2019 November 3), and sector 24 (2020 April 16).  The nova eruption was discovered on 2021 March 18.

{\bf V1369 Cen}~~A closeup look at the $TESS$ sector 11 light curve shows a nearly four-hour periodicity.  The DFT shows only one peak extending significantly above the background noise, and that peak is far above the noise.  A chi-square fit to a sinewave gives the best period of 0.156556 days.  The nova is just off the edge of the images in $TESS$ sector 38, and no confirming data are available from any other source.  I am requiring significant detections in two independent data streams, to avoid most types of artefacts and to avoid random flickering appearing as an apparent DFT peak.  A simple solution is possible when a single source shows a strong periodicity with coverage over many cycles, and that is to split the data set into halves.  For V1369 Cen, I have split the sector 11 light curve at the one-day gap in the middle between orbits.  I find that each half has the identical period, amplitude, and epoch, with the signal being highly significant in each.  Therefore, V1369 Cen has a reliable and accurate period, which can only be orbital.

{\bf FM Cir}~~Schaefer (2021) reported a period of 3.4898 days, based on $TESS$ sectors 11 and 12.  Upon further analysis, this has proven to be an artefact.  In particular, small variations in the background produced broad dips on a time-scale of several days, and one of the standard correction algorithms (the `regression corrector') did a slightly imperfect job at this correction, leaving small but highly-significant dips is the resultant light curve, which then produced a high DFT peak from the coincidences of several of the dip time intervals.  The same collection of spurious uncorrected dips occurred in the standard analysis of both sectors 11 and 12, so that the false periodicity was visible in two data sets.  The problem was recognized in the normal course of triple-checking prior results (after sector 38 data became available), when two other standard correctors did {\it not} show the 3.5 day dips.  With the elimination of this low frequency noise, the DFTs for each of sectors 11, 12, and 38 are all left with just one peak far above the noise and these are all at the same period.  This one single periodicity appears as a stable, significant, and coherent signal from three independent data sets from 2019--2021.  The chi-square fit of a sinewave to all three sector light curves gives $P$=0.1497672 days.  The formal error bar on $P$ in the table is dominated by the possibility that the cycle count from 2019 to 2021 could be 1 larger than for the best cycle count.

{\bf V339 Del}~~I have spent many nights in 2015 and 2016 measuring $V$-band time series on the late fading tail of this bright nova with the Highland Road Park Observatory 20-inch telescope in Baton Rouge, Louisiana.  No significant periodicity was recognized.  With hindsight, I can see the $TESS$ orbital periodicity, at the 6.5-sigma confidence level when taken in isolation, but not at a level that I could call this data set as a confirmation.  The V339 Del $P$ is resolved with the discovery of a highly-significant sinusoidal modulation at 0.162941 days coherently through $TESS$ sectors 14 and 41.

{\bf KT Eri}~~KT Eri is listed in the Ritter \& Kolb catalog as having an orbital period of 0.0938 days, while the VSX catalog lists the period as 737 days.  Both of these are authoritative lists of nova periods , yet they alternatively had KT Eri as either the longest or nearly-the-shortest known nova orbit.  Further {\it published} photometric periodicities are 35.09 s, 56.7, 210, 376, 750, and 752 days.  It turns out that all of these claims come down to observers looking for some limited time, spotting a few ups and downs, then claiming a periodicity based on a relatively few cycles.  All these claims have now been disproven by simply watching KT Eri for a longer time interval and seeing that the claimed periodicities rapidly fail.  The trap is that random flickering or shot noise or fluctuations in CVs will always produce a peak in a DFT that can be confused as a periodicity that appears with a period that is a moderate or large fraction of the observing interval.  This illustrates why any reliable periodicity must be seen for many cycles.  From experience and simulations, more than 8--10 roughly-sinusoidal cycles are required to know the period is reliable.  $TESS$ sector 32 (in 2020 November) shows a light curve with obvious fluctuations that look to be a triangular waveform with a coherent period of 2.64$\pm$0.04 days.  The modulations vary in amplitude from 0.35 mag to 0.15 mag, in a pattern that looks like a beating phenomena.  This is based on just 10 cycles of modulation, so, sector 32 when taken alone, it is on the edge of being reliable.  $TESS$ sector 5 (2018 November) has a much lower amplitude of apparently-chaotic flickering (around the 10 per cent level), all with no significant signal at any period.  The dilemma is that perhaps the orbital period is near 2.64 days with large amplitude changes (like already well-documented for V394 CrA, V2487 Oph, and U Sco; all recurrent novae of similar periods), or perhaps the fluctuations in sector 32 are just rather-unlucky timing of flickers that happen to look like a coherent periodicity.  A convincing solution comes from the many optical spectra recorded with high cadence from 2009--2021 as part of the The Stony Brook / SMARTS Atlas of (mostly) Southern Novae (Walter et al. 2012).  F. Walter finds a significant periodicity in the radial velocity curve for multiple lines with $P$=2.61595 days.  This spectroscopic confirmation of the photometric period produces a reliable orbital period.  Full details on KT Eri appear with exhaustive observations coverage and analysis for the photometric data and for the spectroscopic data (Schaefer et al. 2022b; Walter 2022, in preparation).

{\bf V407 Lup}~~Aydi et al. (2018) claim that V407 Lup is an intermediate polar (IP), with an orbital period of 0.149 days (3.57 hours).  This is based on interpreting the DFT of the short {\it XMM} light curve as showing sidebands of a spin period, where they selected one pair of peaks, with their frequency difference giving the presumed beat period, therefore they deduced that the orbital period would then be 0.149 days.  With this, they then sought and claimed to have found 0.149 day periods in light curves from the {\it Swift} UVOT and {\it XMM} Optical Monitor.  The X-ray `sidebands' are not seen in the {\it Chandra} DFT, and more critically, the frequency spacing of the {\it XMM} `sidebands' corresponds to periods of 4.27, 4.13, 3.61, and 3.31 hours (with uncertainties around $\pm$0.18 hours), with this inconstancy refuting the possibility of these DFT peaks being sidebands and hence that the orbital period is 0.149 day.  Further, the claimed periodicity is not significant in either of their two data sets.  Using the same UVOT data, I find that their 0.149 day DFT peak is just the {\it third} highest noise peak amongst the `grass' of many peaks of similar height, which is to say that this peak is not significant, while the formal significance corresponds to a Gaussian 0.3-sigma confidence level (for the range of test periods from 3--5 hours).  For the optical data from {\it XMM}, it turns out that they only have 1.28 cycles of the alleged 0.149 day periodicity, showing just one minimum and one maximum, both broad and noisy, which is to say that their periodicity is not significant, nor even mildly suggestive of a vague orbit. There is no useable evidence supporting a 0.149 day periodicity (or any periodicity from 3 to 5 hours).  On top of this, there is convincing evidence against anything like a 0.149 day period from other sources.  Aydi et al. point to the lack of any such periodicity in their small selection from the AAVSO data set, while even this small set should show the periodicity for their claimed optical and near-UV amplitudes.  Further, using all of the large AAVSO data set, no significant periodicity is seen from 3--5 hours down to amplitudes of 0.03 mag.  Decisively, the {\it TESS} light curves for sectors 12 and 38 show no significant DFT peaks for periods anywhere from 3--5 hours, with amplitude limits of 0.0010 and 0.0004 mags respectively.  These proofs are that there are no X-ray sidebands, no useable evidence for any period from 3--5 hours, and no such periodicity is seen in three massive data sets to very deep limits.  With the model, the optical spin period, and the orbital period of Aydi et al. being wrong, there remains the question as to the real orbital period of V407 Lup.  For this, Schaefer (2021) gives the orbital period as 3.513 days, although this was pointedly stated to not have high reliability because at the time only one data source ($TESS$ sector 12) was used without any confirmation.  This 3.513 day periodicity is easily visible in the extracted light curve, it produces the highest DFT peak at a highly-significant level, and has been independently confirmed by an analyst from MAST.  For further confirmation from an independent light curve, the AAVSO data have a DFT whose highest non-artefact peak is at 3.6 days, at a formally significant level.  This is actually a double confirmation, as the same DFT peak appears as the highest non-artefact peak for the large independent data sets of  G. Myers (MGW) and B. G. Harris (HMB).  For another confirmation, $TESS$ sector 38 has a 3.7 day periodicity as the highest non-artefact DFT peak.  Note that the $TESS$ periods of 3.5 and 3.7 days are consistent, because ordinary flickering and variations will skew individual maximum and minimum times by a bit, with this leading to substantial uncertainty in the best period for light curves with relatively few cycles.  The 3.5 day periodicity appears as the highest non-artefact DFT peaks in each of four massive independent data sets.  This is conclusive.  Nevertheless, there are fair grounds for worrying about the confidence level of the 3.5 day periodicity.  One worry is that the DFTs for the AAVSO data have issues, with the 3.5 day peak not prominent in the data from 2019, there is confusion from the daily alias peaks, and all the DFTs have a handful of peaks with powers just a bit below that of the 3.5 day peak.  A deeper worry is that a weak $\sim$3.5 day peak appears in DFTs for {\it TESS} light curves for one sector of some nearby stars.  Despite these worries, the fact that {\it four} independent massive data sets all have the 3.5 day period as their highest non-artefact DFT peak makes a very convincing case for a significant, stable, and coherent periodicity.  The next task is to refine the period.  The most accurate $P$ comes from the AAVSO data, with its two year coverage, yielding 3.62$\pm$0.05 days.

{\bf HR Lyr}~~Three $TESS$ sectors (14, 26, and 40) all show a period of 0.905778 days.  All three DFTs have this periodicity as the only peak above the background noise (see Fig. 1), and all three peaks are at the highly-significant level.  The three peaks are at the same period.  The listed period is from an overall chi-square fit to all three sectors.  The folded light curve looks to be a simple sinewave.

{\bf GI Mon}~~$TESS$ sector 7 has a DFT that displays only two peaks above the noise level, both of these are far above the noise level, and the two periods (0.4476 and 0.2239 days) are exactly a factor of two (within uncertainties) of each other.  $TESS$ sector 34 has a DFT that displays only two peaks above the noise level, both of these are far above the noise level, and the two periods (0.4472 and 0.2240 days) are exactly a factor of two (within uncertainties) of each other.  The only reasonable solution is that the true orbital period is the longer of the two, with a deep secondary eclipse providing the DFT power for the signal at half the period.  (The $P$ cannot be the shorter period, as there is no way for the system to remember that its even-numbered minima must be deeper than its odd-numbered minima.)  In sector 7, the DFT power for the longer-period peak is only 27 per cent of the peak power of the shorter period, while in sector 34, the longer period has peak power that is 81 per cent of the power of that of the shorter period peak.  (With deep secondary eclipses, the $P$/2 DFT peak will naturally have more power than the peak for the true orbital period.)  This means that with some variability, the depth of the secondary minimum varies from 50--90 per cent of the depth of the primary minimum.  Indeed, the phase-binned folded light curve (see Fig. 2) displays a classic shape for CVs, with the primary minimum having a depth below the maximum of 0.078 mag, while the secondary has a depth of 0.041 mag.  The quadrature after the primary minimum is significantly brighter than the quadrature before the primary minimum, with the difference being 0.022 mag.  Note, these differential magnitudes are good for knowing the shape of the light curve, however, the $TESS$ photometry aperture (3$\times$3 pixels or 63$\times$63 arcseconds) contains two star substantially brighter than GI Mon, two stars of comparable brightness, plus seven fainter stars visible on the Palomar plates, hence the fluxes in the $TESS$ light curve have an additive constant that makes for greatly lower apparent amplitudes.  A chi-square fit of all the $TESS$ fluxes to this shape light curve gives a period of 0.4470645 days.

{\bf QY Mus}~~The de-trended light curve for $TESS$ sector 11 has the highest DFT peak at a period of 0.912 days, while $TESS$ sector 38 has only one high peak at a period of 0.899 days.  The difference in these periods is not significant, as the observed ordinary flickering will shift the measured maxima and minima, with this being substantial for QY Mus (with the flickering being comparable in size as the periodic modulation) and with only 30 cycles in each sector.  (As an aside, the tail of the point-spread-function of the nova overlaps somewhat with that of a slightly brighter star that displays a highly significant periodicity of 0.745 days.)  The signal from the nova is coherent across the 27 days of each $TESS$ sector, is highly significant in each $TESS$ sector, and stable from 2019 to 2021.  A chi-square fit for all the $TESS$ fluxes to a sinewave gives $P$=0.901135 days.

{\bf V357 Mus}~~This faint nova peaked in January 2018, and has $TESS$ data from sectors 10 and 11 (in the tail of the eruption starting in March 2019) and sectors 37 and 38 (in quiescence starting in April 2021).  All four sectors show a single DFT peak at the same period of 0.155 days.  The sinusoidal modulation amplitude varies by just over a factor of two between the sectors, but the periodicity remains stable and coherent.  The periodicity is highly significant in sectors 10, 11, and 37, with the DFT peaks standing isolated far above the noise.  (Higher noise and lower amplitude in sector 38 makes for a modulation that does not pass my strict significance threshold when taken alone, but given the blatant DFT peaks at the same period and phase as in three independent sectors, this DFT peak shows that V357 Mus was displaying the same modulation even in this last interval.)  The uncertainty for $P$ in Table 1 is from the $\pm$1 uncertainty in the cycle count from 2019 to 2021.

{\bf V390 Nor}~~The only useable light curve for this faint nova is from $TESS$ sector 39.  The DFT shows a singular highly-significant peak at 0.1713 days (see Fig. 1).  The folded light curve is a coherent and stable sinewave.  A very weak first harmonic at 0.0856 days suggests a shallow secondary eclipse, although this is not significantly seen in the folded light curve.  To get my required two independent light curves, the $TESS$ data can be divided into its two orbits, with both orbits displaying highly-significant modulations with identical periods, amplitudes, and epochs.  The $TESS$ photometry aperture is 3$\times$3 pixels or 63$\times$63 arc-seconds and contains four foreground stars substantially brighter than the nova.  The nova's amplitude is certainly larger than the 0.004 mag listed in Table 1.

{\bf V2109 Oph}~~The K2 mission of the $Kepler$ satellite has a wonderful 74-day nearly continuous light curve for V2019 Oph, with 1745-s time resolution.  The nova is faint, each flux in the light curve has typically 3 per cent uncertainty.  The light curve shows 20 per cent sinusoidal modulation for which a periodicity is apparent to the eye, despite having some ordinary flickering.  A DFT reveals a single peak far above the background, with a period of 1.32379 days (see Fig. 1).  To counter a variety of problems with period discovery, I have required that the modulation be detected significantly in two independent data sets.  K2 is the only useable data source for period searches in V2109 Oph.  I have divided the K2 light curve into two segments (at the natural break point of the 3-day gap just before the middle), and these provide my two independent data sets.  The 1.32379 day period is highly significant in both, and coherent across the two parts, hence I take the period as being confirmed.

{\bf V2487 Oph}~~With A. Pagnotta, from 2002--2009, we measured 755 magnitudes of V2487 Oph at the McDonald (McD) and Cerro Tololo (CT) observatories, seeking an orbital period.  As part of our larger study of V2487 Oph, we searched 3760 archival sky patrol photographs from 1890--1989 (Pagnotta \& Schaefer 2014) for eruptions prior to the known 1998 eruption, and indeed discovered the 1901 eruption (Pagnotta et al. 2009), thus making V2487 Oph into the tenth known galactic recurrent nova (Schaefer 2010).  Our photometry does not show any obvious or persistent periodic eclipses or modulation.  When combined with the 1054 magnitudes from the AAVSO, ZTF, Pan-STARRS, and the Palomar Transient Factory, we did find an apparently significant peak at 1.25$\pm$0.03 days (Schaefer, Pagnotta \& Zoppelt 2022a).  Further, we proposed for and received a K2 run with 59-s time resolution (GO 9912, PI Pagnotta), with high-accuracy photometry for a nearly continuous 67 day interval.  A DFT of the K2 data does show a single peak near $P$=1.24$\pm$0.02 that has twice as much power as any other peak.  (Startlingly, V2487 Oph was discovered to produce Superflares, see Schaefer et al. 2022a.  These flares are up to 1.10 mag amplitude and 10$^{39}$ ergs of optical energy, all with an impulsive spike at the start of the light curve with a typical rise time of one minute and a typical duration of around one hour.  These Superflares recur on a time-scale of once a day.  We argue that these extreme Superflares can only be caused by magnetic reconnection events, as seen in Superflare stars, ordinary flare stars, RS CVn stars, and white-light solar flares, including the Carrington Event.  The V2487 Oph Superflares are exciting and have very broad implications.)  This is the second result picking out a period of 1.24 days.  A third method for deriving a 1.24 day orbit is from the spectral energy distribution (SED).  With SED models of the standard $\alpha$-disc, Schaefer et al. (2022a) derive that the orbital period must be between 1.1 and 2.4 days. This is our third measure pointing to $P$$\approx$1.24 days.  With {\it three} independent measures all pointing to $P$=1.24$\pm$0.02 days, I judge the orbital period for V2487 Oph to be reliable.

{\bf V2574 Oph}~~This 2004 nova has good coverage from the K2 mission in late 2016, with 98655 fluxes measured with 58.84-second time resolution for 74.2 days continuously (with only the usual 3-day gap a bit before the middle).  In the range that I search for orbital periods (0.04--10 days), there is only one DFT peak significantly above the noise level, and that has high peak power, in both parts of the light curve (i.e., before and after the gap).  This peak, at 0.135 days, is highly significant, stable and coherent over the 74.2 days of the K2 cycle 11.  For a chi-square fit to a sinewave, $P$ is 0.1350862 days.  The phase-binned folded light curve shows a simple sinusoid.

{\bf V392 Per}~~V392 Per is unique in being a known and monitored dwarf nova system {\it before} its classic nova eruption in 2018.  All three light curves from ZTF (452 magnitudes from 2019.6 to 2021.4), AAVSO (28725 magnitudes from 2019.7 to 2020.7), and $TESS$ sector 19 (1124 fluxes over 25 days starting in 2019 November) display a highly significant DFT peak at the same period.  These DFT peaks are all the highest, far above the background noise level, with no other peaks significantly above the noise level.  (This excludes known artefacts, for example daily aliases of the orbital period and at integer frequencies.)  In all three data sets, the periodicity is coherent, stable, and significant from beginning to end.  The period is consistent across all three data sets, with the AAVSO period being the most accurate due to its length and the number of input magnitudes.  With this, V392 Per has the orbital period of 3.21997 days.

{\bf V400 Per}~~The ZTF light curve of 659 magnitudes from 2018--2021 has four strong peaks in its DFT, at periods of 0.452, 0.826, 1.261, and 4.759 days (see Fig. 1).  Such is often seen for novae in the ZTF data, because the observations are taken from only a restricted range in sidereal time (as for any series of observations from a single earth-based observatory), with all four peaks being simple daily aliases.  The question is then to decide which is the true period.  The ZTF observations are spread out over a range of 0.48 days in sidereal time, and the choice can be made confidently.  The best way is simply to note that the 0.826 day peak is substantially higher than the other aliases.  (Barring cases with substantial noise in the peak power and a narrow range of observed sidereal times, the true period is always represented by the highest DFT peak.)  Further, realistic simulations show that the only way to reproduce the observed relative heights of the alias peaks is for the true period to match the 0.826 day peak.  Further, the DFT of the data with the middle-sidereal-times excluded has the 0.826 day peak emphasized, while the other peaks are lowered.  Further, my analysis of the RMS scatter of magnitude-differences as a function of the phase-differences shows the 0.826 day period to provide the best explanation amongst all the aliases.  This periodicity is stable and coherent over four years, and is highly significant.  A chi-square fit to a sinewave of all the ZTF light curve yields $P$=0.826387 days.

{\bf HS Pup}~~In $TESS$ sectors 7 and 8 (from early 2019) and sector 34 (from early 2021), I found a highly significant periodicity at a period of 0.1786 days.  This DFT peak is far above the background, is the only peak significantly above the background, and has same amplitude and period in all three sectors.  The signal is coherent over the 50 days of sectors 7 and 8.   A chi-square fit to a sinewave of all three sectors gives $P$=0.178641 days.  The error bar quoted in Table 1 is dominated by the uncertainty of 1 in the cycle count from 2019 to 2021.  The folded light curve closely matches a sinewave.

{\bf V598 Pup}~~$TESS$ light curves in sectors 6, 7, 33, and 34 all show only two DFT peaks significantly above the background, with both always being highly significant in each sector.  The two periods are 0.16286 days (with variations of up to 0.00028 days between sectors) and 0.15519 days (with variations of up to 0.00289 days between sectors).  The period variations for the 0.16286 day period are consistent with ordinary flickering shifting the times of maximum light.  The variations of the 0.15519 day modulation is too large for flickering, and is so large that the cycle count could be lost from the start to the end of a sector, which is to say that the modulation is not coherent.  The 0.16286 day period is the stable, coherent, and significant periodicity that must be the orbital period.  A chi-square fit for a sinewave gives $P$=0.162874, while the error bar on $P$ is from an uncertainty of $\pm$1 in the cycle count from 2019 to 2021.    

{\bf YZ Ret}~~This bright nova peaked at 3.7 mag on 2020 July 12.  $TESS$ observed it during five sectors, two sectors are pre-eruption, while three sectors are high on the eruption tail just after the transition.  $TESS$ sector 3 (starting 2018 September 20) and sector 4 (starting 2018 October 19) are ten months before the eruption.  Both pre-eruption sectors show a sinewave visible by eye in the raw data.  Both have only one DFT peak above the background noise level, both are highly significant, and at the same period, so this must be the orbital period.  A chi-square fit of a sinewave to both sectors gives $P$=0.1324539 days.  $TESS$ sectors 29, 30, and 31 run 2020 August 26 to 2020 November 16, from just after transition, when the shell becomes optically thin enough to see the inner binary.  These three $TESS$ sectors in the tail of the eruption do {\it not} show the orbital periodicity.  Rather, a complex set of variations appears near the orbital period, and these constitute two unique and unexplained phenomena (see Section 5). 

{\bf GR Sgr}~~With A. Pagnotta, for GR Sgr, we measured {\it BVR} photometry in 2011 from Cerro Tololo, plus 40 magnitudes in quiescence from the years 1899 and 1923 with the Harvard plates.  I added public-domain data from before 2018 with the Pan-STARRS (2009--2014) and OGLE (2001--2003) experiments, with the OGLE $I$ band magnitudes reported in Mr\'{o}z et al. (2015) and kindly passed along by P. Mr\'{o}z.  Alas, all these pre-2018 magnitudes were not useful for a period search because they are too sparse, scattered over too many years, and all coming from a similar longitude.  For 2018--2021, ZTF provides 392 magnitudes in the {\it zr} and {\it zg} bands.  The light curves in the two bands are normalized to each other by a constant offset, and this procedure is fine for period searches because any colour variations are negligibly small.  From HJD 2458665 to 2458726, the well-sampled light curve of GR Sgr goes through what appears to be two full cycles of modulation with an amplitude of 0.5 mag in amplitude, for an apparent period near 30 days.  A similar periodicity is seen in all the years 2018--2021, but the minimum-to-minimum times vary substantially around any constant period.  The ordinary flickering and fluctuations are superposed on the orbital modulation, making for these variations in peak and minima times.  The ZTF magnitudes were all taken from around the time of culmination from one longitude, and this makes for a difficult daily alias problem.  In particular, this apparent 30 day periodicity also produces DFT peaks with periods near 0.964 and 1.032 days.  P. Mr\'{o}z was kind enough to pass along his further OGLE light curve for 2018--2019, with 105 $I$-band magnitudes.  A DFT of the OGLE data has the same structure, with peaks for periods of 31.5, 1.026, and 0.969 days, with a complex structure of yearly aliases for each peak.  When combining the ZTF and OGLE data, in 2019, four-and-a-half cycles appear with a $P$$\sim$30 day modulation.  The joint DFT reveals the same structure, with complex peaks around 29.4 (or 31.8) days, 1.032 (or 1.035) days, and 0.967 (or 0.969) days, all with similar peak powers.  With the OGLE from Chile and ZTF data from California having sidereal times ranging over only 0.38 days, concentrating on the magnitudes taken from the rising and setting measures does not resolve the aliases.  The absolute magnitude or SED cannot be used to distinguish between the aliases, because the nova is part of a very close blend of three stars, with this confusing and confounding the SED fluxes and the $Gaia$ parallax.  The saving solution is to get just a few magnitudes from a greatly different longitude.  For this, on my request, G. Myers (Past President of the AAVSO) used his remote-control observatory at Siding Springs in Australia to get 22 $V$-band magnitudes.  These magnitudes are now available on the usual AAVSO data download web page.  Now, the joint light curve covers 0.75 days of sidereal time, and it is easy to distinguish that the $\sim$1 day DFT peaks are aliases, with the true period being close to 30 days.  The solution is apparent from the joint AAVSO/ZTF/OGLE DFT, where the periods near one-day have their peak powers substantially lowered, while the one-month period has its peak power raised.  This solution can also be seen in the folded light curves for the various candidate periods around one day, where the new AAVSO magnitudes always greatly disagree with the predicted light curve for either of the $\sim$1 day aliases.  This is a sure solution for the daily alias problem, with the true orbital period being near one month.  The lesser problem remains of deciding between yearly aliases.  The DFTs show peaks near 27.3, 29.5, 31.8, and 34.7 days.  The longest and shortest of these are easily rejected for failing various tests, and for having substantially lower DFT peak powers.  The 29.5 day DFT peak is always substantially higher than the alternative.  The fitted sinewave light curve for the 29.5 day period has the least scatter and the lowest chi-square (by 28) of all the candidate periods.  With these primary indicators both selecting out the 29.5 day peak as the best period, I am taking this to be the true period of GR Sgr.  This periodic modulation is coherent over 2001--2021, and is highly significant in several independent data sets.  This period is best measured from the chi-square fit to all the 2001--2021 data, with $P$=29.4956 days.

{\bf V5558 Sgr}~~The K2 mission has an excellent 69-day continuous coverage of the nova with 1765-second time resolution, except for a 4-day gap just before the middle.  This observation was proposed and accepted for Cycle 9 (GO-9917, PI B. G\"{a}nsicke).  B. G\"{a}nsicke had several years ago discovered the periodicity in V5558 Sgr, but this has never been presented or published.  As part of my systematic period-search through large numbers of novae, I independently discovered the periodicity for V5558 Sgr.  The DFT of the K2 light curve shows just one peak above the noise level, and that peak is extremely high in power.  To meet my requirement of confirmation with two data sets, I have broken the K2 light curve into two parts (at the four-day gap just before the middle), with both parts still displaying a highly significant modulation with the same period, epoch, and amplitude.  The folded light curve shows a simple sinewave.  The full amplitude of the modulation is 0.00256$\pm$0.00010 mag.  However, the photometry aperture is large enough to include several brighter foreground stars, hence the amplitude of the nova alone is substantially larger than in Table 1.  The folded light curve has remarkably little scatter, with the ordinary flickering having an RMS variation of less than 0.0010 mag.  This periodicity is stable and coherent over the entire 69-day interval, with $P$=0.185808 days.

{\bf V697 Sco}~~This nova has the only useable data set as $TESS$ sector 39, covering 28 days in mid-2021 with 18284 flux measures with 120 second time resolution.  The DFT shows a very high peak, far above the background noise (i.e., highly significant) at a period of 0.63 days.  I can get two independent light curves by breaking sector 39 nearly in half (at the one-day gap between orbits), with the same periodicity being very significant in both halves.  The DFT also shows lower peaks at periods of 1.26, 0.42, and 0.316 days, with each of these peaks being far above the background noise level.  This situation is characteristic where the true orbital period is 1.26 days, and the DFT shows peaks at frequencies 2$\times$, 3$\times$, and 4$\times$ the fundamental because the light curve has a prominent secondary eclipse and small asymmetries.  Indeed, a phase-binned and folded light curve shows a deep V-shaped primary eclipse, a shallower U-shaped secondary eclipse, with the egresses being slower than the ingresses for both eclipses, while the two quadrature phases having equal maxima (see Fig. 2).  In flux units, the primary eclipse is 2.3$\times$ deeper than the secondary eclipse.  The amplitude of the primary eclipse is not well measured in magnitude units, because there is substantial uncertainty in the background level (with the primary eclipse going to negative flux in the official SPOC light curve).  The eclipse must be very deep, with a limit of something like $>$1 mag.  A formal chi-square fit to an appropriate eclipsing light curve shape returns $P$=1.26716 days.

{\bf V719 Sco}~~The TESS light curve for sector 39 in mid-2021 has a DFT with only one peak, a highly significant peak, with a stable periodicity of 0.43639$\pm$0.0039 days.  To provide my required two independent light curves that each display the periodicity independently, the two TESS orbits individually display the periodicity significant in both orbits at the same period, phase, and amplitude.  (The same periodicity is seen significantly even in the various eighths of the light curve.)  The photometric aperture contains substantial light from nearby stars, therefore the amplitude of modulation for V719 Sco alone is substantially larger than the quoted 0.0017 mag.

{\bf V1186 Sco}~~For Cycle 11 of the K2 mission, two proposals (GO-11026 PI E. Breedt, GO-11043 PI M. Orio) were accepted for V1186 Sco, resulting in a good light curve for 74-days in 2016 with 1765-s time resolution.  A glance at a blow-up of the light curve shows an obvious periodic eclipse.  This period is highly significant, coherent, and stable.  The folded light curve (See Fig. 2) shows a broad primary eclipse with a depth of 0.182 mag.  The primary eclipse duration is roughly half the period, and there must be some additional mechanism making the V-shaped minima.  The secondary eclipse is relatively short (near 0.22 of the orbit), shallow (0.021 mag), and round-bottomed.  The maximum before the secondary eclipse (at phase 0.41) is the brightest time, while the maximum after the secondary eclipse (at phase 0.63) is fainter by 0.007 mag.  The scatter in the folded light curve is  small, with an RMS of 0.017 mag for the flickering.  A chi-square fit to a realistic light curve template gives the orbital period of 0.202968 days.  $TESS$ sectors 12 and 39 also have highly-significant DFT peaks for V1186 Sco, although with substantially larger noise than for K2, which confirms the period.

{\bf V373 Sct}~~The ZTF light curve has 763 points spread evenly over four year.  The DFT shows three clear peaks, well above other peaks, at periods of 0.450, 0.819, and 4.58 days.  These are all simple daily aliases of each other.  The signal is highly significant and coherent over four years, with no sign of artefacts, hence the period intrinsic to V373 Sct must be one of those three aliases.  The ZTF observations span a range of 0.49 days in sidereal time, and this is enough to allow for a confident identification of the true period.  The simplest way is to note that the 0.819 day peak is substantially higher than the other two aliases, with the height difference corresponding to a strong probability that the peak is the true period.  A further test is to construct a DFT with the middle-range sidereal times times excluded, with the 0.819 day peak becoming even higher relative to its aliases.  Further, realistic simulations of the light curve show that the observed relative heights of the DFT peaks are reproducible only for the input period of 0.819 days.  A sinewave fit with a chi-square analysis gives $P$=0.819099 days.

{\bf XX Tau}~~The period for XX Tau can be solved by $TESS$, with its good time coverage, while the ZTF coverage is also good enough for this task.  $TESS$ sector 32 and the ZTF light curves {\it both} show {\it two} highly-significant roughly-sinewave modulations that match in period and epoch.  Both periodicities have DFT peaks far above the noise, and both are perfectly coherent across the 26 days in late 2020 from $TESS$ sector 32 and across the 982 days of the ZTF data.  The two independent periods are 0.1293567 and 2.929974 days.  One of these two stable and coherent modulations must be the orbital period.  This ambiguity is easily and surely resolved by considering the absolute magnitude of XX Tau.  For a quiescent $V$ magnitude of 18.9 (Vogt et al. 2018), the $E(B-V)$ of 0.22 mag (\"{O}zd\"{o}nmez et al. 2018), and the $Gaia$ parallax of 0.628$\pm$0.227 milli-arcseconds, the absolute magnitude is $+$7.3.  This is normal for a small red dwarf companion plus a low-accretion-rate disc, as requiring a short period.  This absolute magnitude is impossible for a subgiant evolved companion star, as required by a 2.9 day orbital period.  With the ambiguity resolved, the orbital period of XX Tau is 0.1293567 days.  

{\bf V549 Vel}~~$TESS$ covers the late tail of this nova (discovered on 2017 October 17) with four sectors of data, including sectors 8 and 9 (2019 February--March) plus sectors 35 and 36 (2021 February--March).  An obvious photometric modulation of near-ten hours is easily seen by-eye when looking at the light curve.  The waveform often looks triangular (i.e., like a sawtooth), although near-half of the maxima and minima have their sharp points cutoff.  The phase-binned folded light curve is close to a sinewave.  {\it Gaia} shows two 17th-mag very-red foreground stars within one arcsecond of the nova position, which confuses the quiescent brightness level, and makes for the real amplitude for the nova alone to be much larger than the tabulated amplitude in Table 1.  The modulation is stable and coherent over the 50 day intervals for each pair of consecutive $TESS$ sectors.  The period is measured accurately enough that the cycle count across the gap between the pairs of $TESS$ sectors is confidently known.  With the usual chi-square fit to a sinewave across all four $TESS$ sectors, the orbital period is 0.4031692 days.

{\bf NQ Vul}~~$TESS$ covers this old nova during sector 14 (in middle 2019) and sectors 40 and 41 (in middle 2021).  All three sectors reveal a highly significant DFT peak at a period of 0.146256 days.  The folded light curve looks like a simple sinewave, with the noise and flickering a bit larger than the amplitude.  The signal is coherent across the $TESS$ sectors.  The amplitude is stable over all the $TESS$ data.  The best period was found by a chi-square fit to a sinewave, and the data is of top quality so I can keep the cycle count from 2019 to 2021, with $P$=0.1462568 days for NQ Vul.

{\bf PW Vul}~~The ZTF light curve has a highly significant DFT peak at 0.1285753 days, with this sinewave modulation being coherent for each of the years (2018, 2019, 2020, and 2021) and the two filters ($zg$ and $zr$).  A daily alias of comparable peak power is at 0.1476 days, but this possibility is strongly rejected with the $TESS$ data.  The sector 41 $TESS$ light curve from 2021 has a significant DFT peak with the period and epoch matched with the ZTF modulation.  The fitted orbital period of PW Vul is 0.1285753 days.  

\section{Novae With Evolved Companion Stars}

The novae T CrB and RS Oph are widely-known for their red giant companion stars, with their necessarily-large $P$ being measured by long-term radial velocity curves.  In addition, V3890 Sgr has a M8 {\rm III} red giant companion in a 747.6 day orbit determined with a radial velocity curve (Miko{\l}ajewska et al. 2021).  Determination of exact values of $P$ by radial velocity curves requires many and long observing runs, for which only these three high-profile cases have been examined.  Determination of exact values of $P$ by photometric modulation has been stymied in all cases (except for T CrB) by the chaotic `pulsations' (of unknown origin) with all time-scales, such that time-limited light curves will always produce widely-varying apparent periodicity that have nothing to do with the orbit.  Past these three examples, many other nova systems have apparent red giant companions, with orbital periods that must be months-to-years long, for which no $P$ is known or even estimated.  Previously, all long-$P$ novae have been ignored by the compilers and the modelers, but such will skew and blind the demographics.  This can now be corrected.

My program is to (1) systematically construct spectral energy distributions (SEDs) for many novae, (2) recognize the systems with cool luminous blackbodies above the disc flux that prove the presence of an evolved companion star, (3) fit a blackbody to measure the companions' temperatures and luminosities, (4) calculate a blackbody radius for the companion star to get the size of the companion in these Roche lobe overflowing systems, (5) calculate the orbital period from the best estimate stellar masses plus Kepler's Law.  This program is based on the realization that any nova with an evolved companion star must have a dominating blackbody component in the SED, and if a cool blackbody is seen in the SED then the nova must have an evolved companion star that must have a long orbital period.  With the ubiquitous coverage of the 2MASS and $WISE$ surveys, all novae can be positively tested for any evolved companion star, even out past the distance to the galactic center.  With this, I have made a comprehensive survey galactic novae for recognizing systems with an evolved companion star.  For novae with red giant companions, I think that my survey is nearly complete.

For my period search, I need input of the SED, the distance, the extinction, and the stellar masses.  The SED in quiescence has been constructed with optical data from Pan-STARRS, SMARTS, APASS, plus scattered photometry in the literature, while the infrared data comes from 2MASS, SMARTS, and $WISE$, and the brighter systems even have near-ultraviolet photometry from GALEX.  The SEDs usually span 0.48--12 microns, always covering both sides of the Wien peak.   The SED component from the disc is easily recognized and modeled (with standard $\alpha$-disc models), mainly as providing blue light that the cool companion cannot produce, and is negligibly small around the Wien peak in almost all cases.  The nova distances can be measured from the results from $Gaia$, \"{O}zd\"{o}nmez et al. (2018), and Schaefer (2018).  Many of the novae with red giant companions are confidently recognized as being in our galaxy's bulge population, as based on being within $<$12$\degr$ of the galactic center, having extinction-corrected peak magnitudes around V=7.5$\pm$1.4, measured extinctions consistent with the galactic center, and measured parallaxes consistent with the galactic center.  From my own modeling of galactic coordinates for 402 novae, I see that 48 per cent of observed nova are in the bulge population, with 68 per cent of the bulge population within 7.5$\degr$ of the galactic center.  This puts the distance to the bulge novae as 8000 pc with a one-sigma uncertainty of close to $\pm$1000 pc.  The measured extinction values are collected by \"{O}zd\"{o}nmez et al. (2018), and have useful upper limits placed by Schlafly \& Finkbeiner (2011).  The white dwarf masses are measured from radial velocity curves (see the catalog of Ritter \& Kolb), Shara et al. (2018), and various papers of I. Hachisu and M. Kato (see Hachisu \& Kato 2019).  From the results of Shara et al. (2018), the white dwarf mass is 1.20$\pm$0.15 for for P-, O-, C-, and S-class light curves, 1.00$\pm$0.15 for D- and J-class novae, and 0.90$\pm$0.25 for F-class novae, all in units of solar mass, with higher values expected for fast decline rates.  The companion star masses are apparently in the range 0.8--1.0 M$_{\odot}$, with a typical mass ratio to the white dwarf mass of 0.8 or so.  The uncertainties in all these inputs can be propagated forward to give an uncertainty in the derived $P$.  In most cases, the error bars on $P$ are dominated by the distance uncertainty, with the other error bars being small.

A necessary part of this period determination is to test that the observed historical eruption was actually an ordinary thermonuclear runaway nova event, and that the red-giant-bearing system is the nova.  In practice, this is done by looking at the spectra and light curve of the eruptive event (so as to recognize alternatives like symbiotic novae), and by going back to the original nova position plus examination of the nearby stars for spectral lines and flickering.   Let me give three illustrative examples:  First, V733 Sco was originally suggested to be either a nova or a Mira star, as based on the rather limited Leiden data (Plaut 1958).  But the lack of further Mira maxima brightening to 13.5 mag, either from my examination of the many Harvard plates from 1890--1989, or from any of the many modern variable surveys, rules out the Mira possibility definitively.  Moreover, the claimed red giant infrared colours turn out to be based on a bad coordinate in the SIMBAD data base, whereas the correct position from Plaut plus the counterpart identified by Duerbeck (1987) is for a different position, where the counterpart does not have any detected flux in the 2MASS or $WISE$ surveys, and hence there is no evidence pointing towards a red giant companion in this nova system.  Second, V2110 Sgr certainly has a red giant companion, but the eruption light curve is poorly measured in the literature.  I have examined many Harvard plates and find multiple peaks from 1940 to the early 1950s, and the eruption cannot be that of a regular nova, but rather is a symbiotic nova.  Third, V1310 Sgr has been speculated to be a Mira star seen at $R$=13.2 with the Mira variations being mistaken for a nova event.  But this star is not a Mira, as it is nearly constant as seen in many modern surveys.  Critically, I have found Harvard plate MF 22524 to show the nova in eruption at a position roughly 45 arc-sec SW of the $R$=13.2 star.  Further, my 36-magnitude light curve from first brightening to late in the tail shows an ordinary J(390) nova.  The correct position does not have any apparent counterpart to near the Palomar limit, and specifically it has no system with a red giant.

\begin{table*}
	\centering
	\caption{Orbital periods measured from SEDs and the blackbody radii}
	\begin{tabular}{llllllllllll} 
		\hline
		Nova  &  Year  &  LC   & D (pc)  &  $E(B-V)$  &  $M_{\rm WD}^a$ &  $M_{\rm comp}^b$ &  SED$^c$ &  T$_{\rm comp}$ (K) &  $F_0$ (mJy)   &  R$_{\rm comp}$ (R$_{\odot}$)  &  $P (days)$  \\
		\hline
T CrB	&	RN	&	S(6)	&	920	$\pm$	20	&	0.06	$\pm$	0.04	&	1.35	&	1.08	&	GAP2W	&	2830	$\pm$	70	&	6370	$\pm$	400	&	88.6	$\pm$	2.8	&	287	$\pm$	27	\\
V1330 Cyg	&	1970	&	S(217)	&	2900	$\pm$	600	&	0.3	$\pm$	0.1	&	0.91	&	0.73	&	GZP2W	&	6050	$\pm$	220	&	1.0	$\pm$	0.1	&	1.12	$\pm$	0.25	&	0.50	$\pm$	0.17	\\
RS Oph	&	RN	&	P(14)	&	2700	$\pm$	140	&	0.65	$\pm$	0.10	&	1.33	&	1.06	&	GAP2W	&	3630	$\pm$	130	&	2110	$\pm$	16	&	103	$\pm$	7	&	365	$\pm$	52	\\
V794 Oph	&	1939	&	J(220)	&	8000	$\pm$	1000	&	0.9	$\pm$	0.1	&	0.85	&	0.68	&	RZP2	&	4530	$\pm$	210	&	6.3	$\pm$	1	&	12.0	$\pm$	2.1	&	18.0	$\pm$	4.8	\\
V3664 Oph	&	2018	&	...	&	8000	$\pm$	1000	&	0.72	$\pm$	0.2	&	1.05	&	0.84	&	AP2W	&	2100	$\pm$	50	&	450	$\pm$	20	&	320	$\pm$	50	&	2250	$\pm$	500	\\
GK Per	&	1901	&	O(13)	&	430	$\pm$	10	&	0.30	$\pm$	0.04	&	1.22	&	0.98	&	GAP2W	&	3830	$\pm$	100	&	70.1	$\pm$	2.2	&	2.76	$\pm$	0.15	&	1.66	$\pm$	0.14	\\
V392 Per	&	2018	&	P(11)	&	3600	$\pm$	600	&	0.72	$\pm$	0.15	&	1.30	&	1.04	&	GAP2W	&	6100	$\pm$	330	&	7.5	$\pm$	0.2	&	3.7	$\pm$	0.9	&	2.6	$\pm$	0.9	\\
KY Sgr	&	1926	&	S(109)	&	8000	$\pm$	1000	&	1.0	$\pm$	0.5	&	1.20	&	0.96	&	PMV2W	&	2350	$\pm$	40	&	46	$\pm$	9	&	86	$\pm$	12	&	294	$\pm$	61	\\
V1016 Sgr	&	1899	&	S(140)	&	2600	$\pm$	140	&	0.35	$\pm$	0.04	&	1.20	&	0.96	&	AP2W	&	4550	$\pm$	130	&	51.5	$\pm$	3.6	&	11.0	$\pm$	0.9	&	13.4	$\pm$	1.7	\\
V1017 Sgr	&	1919	&	S(130)	&	1200	$\pm$	30	&	0.25	$\pm$	0.10	&	1.20	&	0.96	&	GLSP2W	&	4720	$\pm$	150	&	48	$\pm$	8	&	4.6	$\pm$	0.5	&	3.7	$\pm$	0.6	\\
V1172 Sgr	&	1951	&	...	&	8000	$\pm$	1000	&	0.4	$\pm$	0.1	&	1.10	&	0.88	&	SW	&	2310	$\pm$	70	&	34.1	$\pm$	3.9	&	76.5	$\pm$	12.8	&	255	$\pm$	64	\\
V3645 Sgr	&	1970	&	...	&	8000	$\pm$	1000	&	0.39	$\pm$	0.03	&	1.10	&	0.88	&	AP2W	&	3490	$\pm$	110	&	9.8	$\pm$	0.5	&	22.0	$\pm$	3.1	&	39.5	$\pm$	8.4	\\
V3890 Sgr	&	RN	&	S(14)	&	8000	$\pm$	1000	&	0.59	$\pm$	0.1	&	1.38	&	1.05	&	ACP2W	&	2230	$\pm$	60	&	260	$\pm$	12	&	223	$\pm$	30	&	1220	$\pm$	250	\\
V5580 Sgr	&	2008	&	...	&	8000	$\pm$	1000	&	0.34	$\pm$	0.1	&	1.10	&	0.88	&	AP2W	&	3900	$\pm$	280	&	11	$\pm$	1	&	20	$\pm$	4	&	34	$\pm$	9	\\
V5581 Sgr	&	2009	&	...	&	8000	$\pm$	1000	&	1.5	$\pm$	0.3	&	1.10	&	0.88	&	SP2W	&	2250	$\pm$	60	&	380	$\pm$	120	&	266	$\pm$	58	&	1660	$\pm$	530	\\
V723 Sco	&	1952	&	S(23)	&	8000	$\pm$	1000	&	0.57	$\pm$	0.10	&	1.38	&	1.10	&	MV	&	3840	$\pm$	190	&	1.6	$\pm$	0.2	&	7.6	$\pm$	1.2	&	7.2	$\pm$	1.8	\\
V745 Sco	&	RN	&	P(9)	&	8000	$\pm$	1000	&	1.0	$\pm$	0.2	&	1.39	&	1.11	&	GS2W	&	2020	$\pm$	60	&	540	$\pm$	20	&	370	$\pm$	50	&	2440	$\pm$	500	\\
V977 Sco	&	1989	&	...	&	8000	$\pm$	1000	&	0.92	$\pm$	0.10	&	1.10	&	0.88	&	MV	&	3740	$\pm$	240	&	6.9	$\pm$	0.8	&	16.8	$\pm$	3.2	&	26.2	$\pm$	7.4	\\
V1313 Sco	&	2011	&	S(18)	&	3100	$\pm$	1600	&	1.00	$\pm$	0.2	&	1.20	&	0.96	&	GS2W	&	3140	$\pm$	90	&	70	$\pm$	2	&	27	$\pm$	13	&	50	$\pm$	35	\\
V1534 Sco	&	2014	&	S(9)	&	8000	$\pm$	1000	&	0.92	$\pm$	0.10	&	1.37	&	1.10	&	S2W	&	2420	$\pm$	80	&	118	$\pm$	8	&	133	$\pm$	21	&	520	$\pm$	120	\\
V1535 Sco	&	2015	&	S(20)	&	8000	$\pm$	1000	&	0.8	$\pm$	0.2	&	0.85	&	0.68	&	GSA2W	&	3610	$\pm$	170	&	12.5	$\pm$	2.2	&	23.7	$\pm$	4.7	&	50	$\pm$	14	\\
V1657 Sco	&	2017	&	...	&	8000	$\pm$	1000	&	1.0	$\pm$	0.1	&	1.10	&	0.88	&	S2W	&	2870	$\pm$	90	&	226	$\pm$	26	&	142	$\pm$	21	&	650	$\pm$	145	\\
EU Sct	&	1949	&	S(42)	&	5100	$\pm$	1500	&	0.84	$\pm$	0.10	&	1.20	&	0.96	&	SP2W	&	3150	$\pm$	110	&	39	$\pm$	3	&	32.7	$\pm$	9.7	&	68	$\pm$	30	\\
FS Sct	&	1952	&	J(86)	&	3600	$\pm$	1500	&	0.5	$\pm$	0.1	&	1.00	&	0.80	&	PR2W	&	3170	$\pm$	190	&	2.6	$\pm$	0.2	&	5.9	$\pm$	2.6	&	5.7	$\pm$	3.7	\\
X Ser	&	1903	&	S(730)	&	5100	$\pm$	2000	&	0.2	$\pm$	0.1	&	1.05	&	0.84	&	GAP2W	&	5380	$\pm$	320	&	1.09	$\pm$	0.03	&	2.45	$\pm$	1.01	&	1.5	$\pm$	0.9	\\
		\hline
	\end{tabular}
	
\begin{flushleft}	
\
$^a$The white dwarf mass is in units of M$_{\odot}$.  The error bars are near 0.03 M$_{\odot}$ for the RNe, and estimated as 0.10--0.20 M$_{\odot}$ for the other novae.  The size of these error bars are negligible for the uncertainty in $P$. \\
$^b$The companion mass is in units of M$_{\odot}$.  The error bars are always poorly known, estimated to be 0.10--0.20 M$_{\odot}$, and negligible for contributing to the uncertainty in $P$.\\
$^c$The sources for the SEDs are indicated with a letter or number:  A--AAVSO (including APASS), C--Schaefer (2010), G--$Galex$, L--Landolt (2016), M--Mr\'{o}z et al. (2015), P--Pan-STARRS, R--Ringwald et al. (1996), S--SMARTS, V--Saito et al. (2013), W--$Wise$, Z--Szkody (1994), 2--2MASS. \\

\end{flushleft}

\end{table*}

I have searched many novae, and I find 25 for which their SED has a clear blackbody component far above the disc light.  These necessarily have a cool and large companion star, and a correspondingly large orbital period.  That is, the mere existence of a highly significant blackbody with 2100--6500 K temperature in the nova system means that we reliably have a very long orbital period.  Out of these 25 novae, 6 have previously known $P$ from radial velocity curves or photometric modulations, 1 has my new photometric period, 1 has a previously known red giant companion, 7 more have previously been suggested to have red giant companions based on IR colors, and 10 novae are newly identified here as having evolved companion stars as based on their SEDs.  In all, this SED analysis produced 18 new $P$ measures of moderate accuracy.

The results from my SED fits and period measures are given in Table 2.  The first three columns are the GCVS name of the nova (in the GCVS order), the year of the nova, and the light curve class (see Strope, Schaefer, \& Henden 2010).  The next four columns give the adopted distance, $E(B-V)$, white dwarf mass, and companion star mass.  Column 8 gives the sources for the SED curve, keyed by initials in the footnote.  Columns 9 and 10 give the results from the SED fit, the effective surface temperature of the companion and the extinction corrected flux at the blackbody peak in milliJanskys.  The last two columns give the derived radius for the companion star (in units of solar radii) and the orbital period (in units of days).

My measurements of $P$ use only standard input and standard physics, and it is a real measure of $P$.  These measured periods are just as reliable as the periods from photometric DFTs and radial velocity curves.  However, a stark difference is that the real error bars on $P$ are not parts-per-million or parts-per-thousand, but rather are up to 30 per cent or so.  My new $P$ measures are {\it reliable}, even if not of high accuracy.  For most purposes, the moderate accuracy of my new method is perfectly adequate.  For example, for demographic and evolution purposes, a 30 per cent uncertainty does not shift the bin in the period histograms.  Even for modeling of individual systems, a 30 per cent error bar in $P$ is of small import, especially when the alternative is to either have no information on the period or only to use the red giant nature of the companion to limit the period to be from 30--2000 days.  My new $P$ measures are reliable and of adequate accuracy for most purposes.

This method for measuring $P$ has substantial uncertainties just due to the usual error bars on the input, with the median fractional uncertainty being 24 per cent for the novae in Table 2.  Further, systematic uncertainties arise from the ordinary variability in quiescence leading to a distorted SED from segments observed at different brightness levels.  A nice test for the overall accuracy of the method is by comparing the SED $P$ versus the $P$ from radial velocity curves or photometric modulation.  Table 2 gives the SED $P$ for seven novae with accurate and reliable periods.  The per cent errors are $+$26 for T CrB, $-$19 for RS Oph, $-$17 for GK Per, $-$19 for V392 Per, $-$36 for V1017 Sgr, $+$63 for V3890 Sgr, and $+$1 for X Ser.  The average per cent error is zero, indicating that the method has no apparent bias high-or-low.  The RMS of the errors is 34 per cent.  This is a horrifyingly large error bar for those of us who commonly measure and use periods at the part-per-million level, but a 34 per cent uncertainty is actually of adequate accuracy for many applications.

Examples of my SEDs have already been listed and displayed in Schaefer (2010) for all ten RNe (including T CrB, RS Oph, V3890 Sgr, and V745 Sco) and in Salazar et al. (2017) for V1017 Sgr.  Further examples are shown in Fig. 3, with the observed fluxes as sourced above, plus the model fits as listed in Table 2.  The SEDs for GK Per, V392 Per, and X Ser are selected to round out the novae in Table 2 for which we have ground-truth periods (see previous paragraph).  GK Per has one of the best SEDs and model fits for those in Table 2.  The SEDs for V392 Per and X Ser are both typical of the SEDs and the model fits.  V392 Per shows the ordinary problem of constructing an SED due to combining measures at different epochs, with the quiescent levels fluctuating up-and-down.  The formal measurement errors on the SED points are always smaller than the plot symbol, and the real dominant error due to source variability cannot be known other than by the scatter around some best fitting smooth curve.  The accretion disc light is modeled from a full integration of $\alpha$-disc models, and is always small outside of the ultraviolet.  The fourth panel in Fig. 3 is for V723 Sco, which is my poorest SED.  Even for this poorest SED, the blackbody shape and temperature are confident.

\begin{figure}
	\includegraphics[width=1.01\columnwidth]{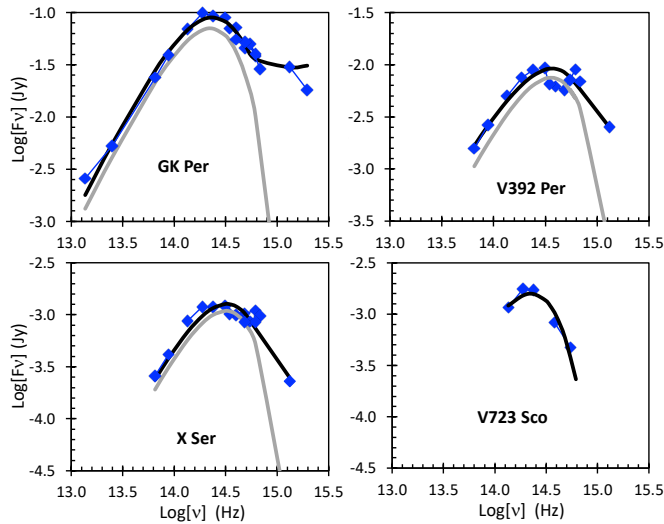}
    \caption{SEDs for four novae with evolved companions.  The best fitting models (thick black curve) are the sum of a blackbody for the companion star (thick grey curve) and an $\alpha$-disc model for the accretion disc.  The SED for GK Per is one of the best for the novae in Table 2, while the SEDs for V392 Per and X Ser are typical examples.  The SED for V723 Sco is the poorest of the SEDs from Table 2, yet the blackbody shape and its temperature are clear.}
\end{figure}

This section is providing a nearly exhaustive survey of novae with red giant companions, and the SED can also identify many novae with subgiant companions.  The distinction between red giants and subgiants is formally based on either of two definitions involving the evolutionary state of the stellar core or by some luminosity range.  These usual definitions are hard to apply to the novae.  Instead I am making an empirical distinction that subgiants are those inside nova binaries with periods from 0.6 to 10 days.  The labels for these systems are not important, but rather the fact that the companions are evolved off the main sequence, as this is what makes for an entirely separate evolution from novae with nearly-main-sequence companions.

I am struck that 15 out of the 20 systems with red giant companions are in the galactic bulge population.  These have an average angular distance from the centre of 8.5$\degr$, while the distances are consistent with 8000$\pm$1000 pc.  This bulge fraction of 75 per cent is to be compared to 32 per cent for novae with subgiant companions and 11 per cent for novae with main sequence companions, for my list of 156 novae with periods.  The bulge fraction of the novae with red giant companions (75 per cent) can also be compared to the bulge fraction of discovered nova at 43 per cent (Hatano et al. 1997).  The three mild selection effects I can identify actually work against the high fraction of red giant novae in the bulge.  Therefore, some unknown mechanism makes the novae with red giants significantly more frequent in the older bulge population.

\section{Confirmed and Improved Orbital Periods}

In searching all available data on various novae, I have confirmed and improved previously claimed $P$ values.  Some of these confirmations are for claims that were in various ways either questionable or ambiguous.  Others are simply improving on prior published periods.  These confirmations and improvements are for V368 Aql, V500 Aql, RS Car, V705 Cas, V2275 Cyg, V2467 Cyg, V972 Oph, V2860 Ori, V2572 Sgr, and V382 Vel.  These confirmations and improvements were all made with $TESS$, $Kepler$, and ZTF light curves (see Table 3), with the high cadences and the long time spans making for significant improvements over the prior published $P$ values.  As confirmations, these periods are reliable because they now have two or more independent measures.

In addition, from my prior papers, I have century-long $O-C$ curves for six classical novae (QZ Aur, HR Del, DQ Her, BT Mon, RR Pic, and V1017 Sgr) plus for six recurrent novae (CI Aql, V394 CrA, T CrB, IM Nor, T Pyx, and U Sco), with these giving very accurate $P$ values plus the first very-long-term (i.e., evolutionary) steady period changes ($\dot{P}$), plus sudden period changes ($\Delta P$) across 12 nova eruptions.  

Table 3 presents the 22 new confirmations and improved periods.  The first three columns describe the nova, with the GCVS variable star name (listed in the correct order from the GCVS), the year of eruption (with RN for recurrent novae with 2--12 known years of eruption), and the light curve class (from Strope, Schaefer, \& Henden 2014).  The fourth column gives the new/improved period in units of days.  The last column cites the appropriate references, keyed to a listing in the table footnote.  %Further, detailed discussion for RS Car and V382 Vel are presented in the next Section.

\begin{table}
	\centering
	\caption{Confirmed and improved periods for novae}
	\begin{tabular}{lllll} 
		\hline
		Nova  &  Year  &  LC class  &  $P$ (days)  &  Ref.  \\
		\hline
CI Aql	&	RN	&	P(32)	&	0.61836092	&	2	\\
V368 Aql	&	1936	&	S(42)	&	0.6905093	&	1 (ZTF), 3	\\
V500 Aql	&	1943	&	S(43)	&	0.145259	&	1 (ZTF), 4	\\
QZ Aur	&	1964	&	S(25)	&	0.35749703	&	5	\\
RS Car	&	1895	&	J(80)	&	0.082436	&	1 ($TESS$), 6, 7	\\
V705 Cas	&	1993	&	D(67)	&	0.228284	&	1 ($TESS$), 8	\\
V394 CrA	&	RN	&	P(5)	&	1.515682	&	9	\\
T CrB	&	RN	&	S(6)	&	227.532	&	1 (AAVSO), 10	\\
V2275 Cyg	&	2001	&	S(8)	&	0.32596	&	1 ($TESS$), 21	\\
V2467 Cyg	&	2007	&	O(20)	&	0.153789	&	1 ($TESS$), 11	\\
HR Del	&	1967	&	J(231)	&	0.21416215	&	12	\\
DQ Her	&	1934	&	D(100)	&	0.1936208997	&	13	\\
BT Mon	&	1939	&	F(182)	&	0.33381490	&	13	\\
IM Nor      	&	RN	&	P(80)	&	0.207165513	&	1 ($TESS$), 14	\\
V972 Oph	&	1957	&	S(176)	&	0.279641	&	1 (K2), 15	\\
V2860 Ori	&	2019	&	S(15)	&	0.422579	&	1 (ZTF), 16	\\
RR Pic	&	1925	&	J(122)	&	0.1450237620	&	12	\\
T Pyx	&	RN	&	P(62)	&	0.07623361	&	1 ($TESS$), 17	\\
V1017 Sgr	&	1919	&	S(130)	&	5.786290	&	18	\\
V2572 Sgr	&	1969	&	P(44)	&	0.157038	&	1 ($TESS$), 7	\\
U Sco	&	RN	&	PP(3)	&	1.23055183	&	2	\\
V382 Vel	&	1999	&	S(13)	&	0.1461	&	1 ($TESS$), 19, 20	\\
		\hline
	\end{tabular}
\begin{flushleft}
References:  1. This paper, with the data source in parentheses;  2. Schaefer 2011;  3. Marin \& Shafter 2009;  \\4. Haefner 1999;  5. Schaefer et al. 2019;  6. Woudt \& Warner 2002;  \\7. Fuentes-Morales et al. 2021; 8.Retter \& Leibowitz 1995; 
9. Schaefer 2009;  10. Leibowitz et al. 1997;  11. Swierczynski et al. 2010;  12. Schaefer 2020b;  13. Schaefer 2020a;  14. Patterson et al. 2022; 15. Tappert et al. 2013; 16. Denisenko 2019;  
17. Patterson et al. 2017;  18. Salazar et al. 2017;  \\19. Balman et al. 2006; 20. Bos et al. 2001; 21. Balman et al. 2005.
\end{flushleft}

\end{table}

\section{Non-Orbital Periods}

In the course of my search for orbital periods, I have found many coherent periodicities that are {\it not} orbital.  Some have already been discussed, as asides, in prior Sections.  These scattered discussions do not show the commonness and commonality of the cases.  Let me collect and highlight these new non-orbital periods.  Table 4 presents the usual nova descriptors in the first three columns (the GCVS name, the eruption year, and the light curve class), then two columns giving the orbital and non-orbital period (both in units of days), while the last column gives whether the non-orbital period is stable or transient.

{\bf RS Car}~~Woudt \& Warner (2002) report a period of 0.08238 days, Fuentes-Morales et al. (2021) report periods of 0.082429 and 0.089842 days (preferring the second daily alias).  With its lack of daily alias problems, the $TESS$ data for sectors 10, 11, 37, and 38 solve this question.  The period is easily visible in all four sectors, with $P$=0.082436 days.  With this being coherent, stable, and significant, and seen in six independent data sets, this must be the orbital period.  However, the $TESS$ sector 10 and 11 light curves also display a coherent and significant periodicity at 0.02788 days (2409 seconds).  This is too short for any orbit.  And this does not appear in sectors 37 or 38, and hence the signal is transient.  A reasonable explanation is that RS Car is an IP with a white dwarf spin period of 2409 seconds.

{\bf V842 Cen}~~Woudt et al. (2009) report {\it seven} optical periodicities; a coherent 56.825 s modulation from the white dwarf spin, sidebands at 56.598 and 57.054 s, an unobserved orbital period of 0.164 days based on a theoretical model, a quasi-periodic oscillation spanning 350--1500 s, a ``strong brightness modulation" at 0.1575 days, and an ``even stronger signal" at 0.12025 days.  {\it None} of these was seen in either sectors 11 or 38 of $TESS$.  In particular, sector 38 had 20 s time resolution, hence the spin period, its sidebands, and the QPO {\it should} have been easily detected.  Further, the theorized orbital period is completely invisible down to the very deep limits from $TESS$, making it unlikely that this theory is correct.  The most likely reconciliation is that the periodicities are transient, and that all six observed modulations were in an on-state during the one month of observations in 2008.  However, the $TESS$ light curves did reveal yet another periodicity.  For sector 38, a sinewave with period 0.1481 days is highly significant, and is coherent from the first to last of the $TESS$ light curve for that 27 day interval.  The 0.1481 day signal appears significantly and coherently in both the first and second halves of the sector 38 light curve, and this period is real and intrinsic to the nova.  However, this confident periodicity does {\it not} appear in $TESS$ sector 11, nor, presumably, in the 2008 fast photometry of Woudt et al. (2009).  The three observed transient periodicities inside the usual range of nova orbital periods are 0.12025, 0.1481, and 0.1575 days.  Any one of these  might be the orbital period, but their detection in only one data set each does not inspire confidence.  Further, the chances of selecting the true orbital period from amongst these three alternatives is only 1-in-3 at best.  This is proof that at least one nova system displays multiple transient periodicities that are not the orbital period.  Further, this proves a clear violation of simple theory, where the apparent spin period and its sidebands do {\it not} have an orbital period at 0.164 days to beat with.

{\bf V2574 Oph}~~V2574 Cyg shows highly significant, stable, and coherent periodicities at 196.141 and 220.659 seconds.  These two signals are not any type of alias or artefact that I can recognize, nor with any apparent relation to the orbital period.  The relative peak powers in the DFT varies somewhat through the K2 cycle, with the two peaks usually of comparable power, although the 220.659 seconds peak is generally higher.  Given the stability of the two signals, one or both might be tied to the white dwarf spin period, perhaps like an intermediate polar.  But the fast periods are not sideband structures, because their frequency spacing is 6.61$\times$ the frequency of the orbit.  I know of no precedent for having exactly-two fast non-orbital periodicities, nearly equal in DFT power, with a wide frequency spacing unrelated to the orbit.  The two fast periodicities are a mystery.

{\bf V407 Lup}~~The AAVSO dataset has 14344 magnitudes (from G. Myers) that displays a highly significant optical periodicity at 0.0068434566$\pm$0.0000000026 days (591.2746 seconds) that is stable and coherent from 2017.7 to 2019.6.  V407 Lup is likely an IP, with the primary evidence being an apparent spin period of 565.04$\pm$0.33 seconds, as seen in the DFT of the {\it Chandra} X-ray light curve.  This DFT peak has no sidebands.  The comparable {\it XMM} X-ray light curve has a prominent sideband structure with the highest peak at 543.3 s, and lower peaks at 563.9, 524.6, 589.8, and 620.4 s (all with uncertainties of near $\pm$0.7 s), in order of DFT power.  Aydi et al. (2018) interprets this as an IP sideband structure with a spin period near 565 s and an orbital period of 0.149 days.  But this simple idea cannot be correct.  (1) The frequency spacing of the `sidebands' is not uniform, whereas the simple model requires that the spacing equals exactly the orbital frequency.  Instead, the `sideband' frequency spacing gives derived $P$ value are 4.27, 4.13, 3.61, and 3.31 hours, with error bars of $\pm$0.18 hours, which is not consistent with a constant.  The 'sidebands' are not beating between the spin and orbital periods, as proven by the non-uniform spacing.  (2) The {\it Chandra} DFT has only one peak at 565.04 s, the optical DFT has the only peak at 591.2746 s, while the {\it XMM} DFT has the highest peak at 543.3 s, with such violating the theory and experience that the spin period corresponds to the highest peak.  (3) The theory-required orbital period (either at 0.149 days or any from 3--5 hours) is not significantly present in any data set (see Section 2.3), where the limits go very deep and any such period must be visible.  (4) I find a coherent periodicity at 3.62 days in {\it four} independent data sets, and that must be the true orbital period, in which case the so-called `sidebands' are nothing of the kind.  With the collapse of the sideband model, the nature of the 591.2746 s optical period remains a mystery.

{\bf QY Mus}~~A non-orbital periodicity is visible in  $TESS$ sector 11, with 77 per cent of the power as for the orbital period, with this highly significant and stable over at least 44 cycles.  This period does not appear in sector 38.  This period (0.6083 days) has no apparent relation to the long-lasting orbital period (0.901135 days).  I know of no mechanism for this transient period, which remains a mystery.

{\bf V598 Pup}~~In sectors 6, 7, 33, and 34 of $TESS$ data, a highly significant modulation appears near a period of 0.155 days.  This periodicity has a substantially higher DFT peak power than does the orbital period for three of the four sectors.  The period varies by up to 1.9 per cent from sector to sector, and the overall Fourier transform shows a greatly broadened peak, and hence this periodicity is not coherent.  The signal varies in amplitude, for example in 2020--1, it peaks in power around BJD 2459211 (roughly one third of the way through sector 33), then starts fading until it has apparently zero power around BJD 2459238 (roughly one third of the way through sector 34), then reappears near the end of sector 34.  The transient period is 4.7 per cent smaller than the orbital period, and some strange negative superhump mechanism is a possibility, but the photometry does not show anything like a superoutburst.  In all, I have no understanding of how a nova system can produce such a transient and incoherent periodicity.

{\bf YZ Ret}~~Starting soon after the transition in the eruption light curve, YZ Ret displayed two new unexpected and unprecedented phenomena.   The first phenomenon is that the brightness showed a `chirping decrescendo', which is to say that a photometric modulation decreased in amplitude down to zero, while the frequency increased.  For the first 15 days of sector 29, the light curve displayed aperiodic dips with amplitudes decreasing from 0.1 mag to under 0.003 mag, all while the minimum-to-minimum times ran from 1.0 days down to 0.3 days.  The second new phenomenon is the two transient periodicities, both close to the orbital period, that come-and-go during the three months of $TESS$ observations in sectors 29, 30, and 31.  YZ Ret shows a 0.1384 day periodicity (4.5 per cent longer than $P$), with this being visible from BJD 2459088 to 2459114 and from 2459158 to 2459169.  The nova also shows a significant periodicity at 0.13393 days (1.1 per cent longer than $P$), with this present from BJD 2459130 to 2459169.  With the transient periods being 4.5 and 1.1 per cent longer than $P$, my speculation is that the phenomenon is something like a superhump, where the disc conditions during the decline of the eruption make for an eccentric precessing accretion disc.  Presumably, with the chaotic conditions just after transition, the delicate conditions required for the superhump mechanism will come-and-go.

{\bf XX Tau}~~This nova shows two periodicities with comparable DFT power, an orbital $P$ at 0.1293567 days, and a perplexing non-orbital period at 2.929974 days.  The non-orbital period is highly-significant, coherent, and stable over the ZTF time interval from 2018.59 to 2021.28.  The mechanism for the 2.929974 day period is a mystery.

{\bf PW Vul}~~Hacke (1987a; 1987b) reports a period of 0.21372 days, and the folded light curve looks significant.  This periodicity does not appear in either the $TESS$ or ZTF light curves, hence the signal must be transient.  The second transient periodicity is that all the portions of the ZTF light curve show a 4.071 day modulation.  (This might be visible in the $TESS$ light curve, but systematic uncertainties in the light curve corrections make this uncertain.)  These two coherent and transient non-orbital periodicities remain a mystery .

\begin{table}
	\centering
	\caption{Coherent Non-orbital Periods in Novae}
	\begin{tabular}{llllll} 
		\hline
		Nova  &  Year  &  LC  &  $P$ (days)  &  $P_{\rm non-orbital}$(d) & Stable?  \\
		\hline
RS Car	&	1895	&	J(80)	&	0.082436	&	0.02788	&	Transient	\\
V842 Cen	&	1986	&	D(40)	&	unknown	&	0.12025	&	Transient	\\
V842 Cen	&	1986	&	D(40)	&	unknown	&	0.1481	&	Transient	\\
V842 Cen	&	1986	&	D(40)	&	unknown	&	0.1575	&	Transient	\\
V2574 Cen	&	2004	&	S(41)	&	0.1350862	&	0.0022701	&	Stable	\\
V2574 Cen	&	2004	&	S(41)	&	0.1350862	&	0.0025539	&	Stable	\\
V407 Lup	&	2016	&	S(8)	&	3.62	&	0.006843457	&	Stable	\\
QY Mus	&	2008	&	S(95)	&	0.901135	&	0.6083	&	Transient	\\
V598 Pup	&	2007	&	...	&	0.162874	&	0.155	&	Transient	\\
YZ Ret	&	2020	&	P(22)	&	0.1324539	&	0.1339	&	Transient	\\
YZ Ret	&	2020	&	P(22)	&	0.1324539	&	0.1384	&	Transient	\\
XX Tau	&	1927	&	D(43)	&	0.1293567	&	2.929974	&	Stable	\\
PW Vul	&	1984	&	J(116)	&	0.1285753	&	0.21372	&	Transient	\\
PW Vul	&	1984	&	J(116)	&	0.1285753	&	4.071	&	Transient	\\
		\hline
	\end{tabular}
\end{table}

\section{Novae Catalogued With Unreliable Periods}

The literature contains many claims for orbital periods of novae that are certainly wrong.  The highly-useful catalogs of Duerbeck (1987) and CV-Cat (Downes et al. 2001) both make a point of identifying dubious period claims.  Mr\'{o}z et al. (2015) correctly reported eight "Post-novae Showing Semi-regular Variability", and it was only later authors and compilers who started to label these as orbital periods.  The grand compilation of nova periods by Fuentes-Morales et al. (2021) did a good job of explicitly separating out the dubious periods and specifying the problems.

For my sample of 31 new, reliable, and accurate $P$ values (see Table 1), 17 of the novae had prior published and cataloged wrong-$P$ values, with an average of 2.0 false-periods per nova.  The large catalogs containing nova periods (VSX, GCVS, the Ritter \& Kolb catalog, and CV-Cat) have error rates on reported periods of 25, 52, 26, and 11 per cent respectively.

With so many false-periods published, I can do no better than to follow Fuentes-Morales et al. (2021), and tabulate all the rejected catalogued $P$s, giving references and specifying the reasons for the rejection.  This is given in Table 5.  The first three columns give the identifying properties of the nova; the GCVS name, the year of eruption, and the light curve class as defined in Strope, Schaefer, and Henden (2010).  The next two columns give the claimed periods and the references.  The last column gives the reason why the claimed period is wrong, with the evidence keyed to letters in the footnotes for each reason.  This list is of 24 novae.  In addition, there are 17 more novae for which my new reliable $P$ measures serve as adequate evidence for rejection of 34 prior claims.

\begin{table}
	\centering
	\caption{Novae with claimed $P$ that are not reliable}
	\begin{tabular}{llllll} 
		\hline
		Nova & Year & LC   &   $P$ claim (days)   &   Ref.   &   Evidence   \\
		\hline
V1391 Cas	&	2020	&	D(119)	&	0.15848	&	1	&	H, I, J	\\
V842 Cen	&	1986	&	D(48)	&	0.164	&	2	&	B, F, G, J	\\
V1047 Cen	&	2005	&	S(20)	&	0.361, 8.36	&	3, 4	&	A, B, F, J	\\
AP Cru	&	1935	&	...	&	0.2133, 0.0213	&	5	&	B, D, J	\\
V2274 Cyg	&	2001	&	D(33)	&	0.30	&	6	&	C, E	\\
V2362 Cyg	&	2006	&	C(246)	&	0.0658, 0.207	&	7, 8	&	C, E, G, I	\\
V2491 Cyg	&	2008	&	C(16)	&	0.71, 0.0958	&	9, 10	&	A, G, I, J	\\
V2891 Cyg	&	2019	&	J(182)	&	0.16148	&	11	&	B, F, J	\\
DM Gem	&	1903	&	P(22)	&	0.1228, 0.0157	&	12	&	A, B, J	\\
DI Lac	&	1910	&	S(39)	&	0.5324	&	13	&	B, C, G, J	\\
DK Lac	&	1950	&	J(202)	&	0.1296	&	14	&	A, H, J	\\
LZ Mus      	&	1998	&	P(12)	&	0.1693	&	15	&	B, C, I, J	\\
V1112 Per	&	2020	&	D(33)	&	0.0927, 0.608	&	16, 17	&	F, G, I, J	\\
V445 Pup    	&	2000	&	D(240)	&	0.650654	&	18	&	A, B, H	\\
V574 Pup	&	2004	&	S(33)	&	0.0472	&	19	&	B, H, J	\\
V1016 Sgr	&	1899	&	S(140)	&	0.07579635	&	22	&	F, M	\\
V4077 Sgr	&	1982	&	...	&	0.16	&	20	&	C, E	\\
V4643 Sgr	&	2001	&	S(6)	&	32	&	21	&	H, K	\\
V5581 Sgr	&	2009	&	S($\sim$10)	&	62.3	&	21	&	K	\\
V5583 Sgr	&	2009	&	S(9)	&	7.101	&	22	&	L	\\
V5980 Sgr	&	2010	&	S($\lesssim$82)	&	0.6332, 1.266	&	21	&	B, J	\\
V745 Sco	&	RN	&	P(9)	&	136.5, 77, 510	&	21, 23	&	A, G, K	\\
V1187 Sco	&	2004	&	S(17)	&	354	&	21	&	K	\\
V1324 Sco	&	2012	&	D(30)	&	0.067, 0.133	&	24	&	B, C, J	\\
V1534 Sco	&	2014	&	S(11)	&	0.61075	&	22	&	L	\\
		\hline
	\end{tabular}
\begin{flushleft}
References:  1. Schmidt 2021;  2. Woudt et al. 2009;  3. Aydi et al. 2021;  \\4.  Aydi et al. 2019;  5. Woudt \& Warner 2002;  6. Ritter \& Kolb 2003;  \\7. Goranskij et al. 2006;  8. Balman et al. 2009;  9.  Baklanov et al. 2008;  \\10.  Zemko et al. 2018;  
11. Schmidt 2020;  12. Rodriguez-Gil \& Torres 2005;  13. Goransky et al. 1997;  14. Katysheva \& Shugarov 2007;  15. Retter et al. 1999;  16. Schmidt 2021;  17. Thomas et al. 2021;  18. Goranskij et al. 2010;  19. Walter et al. 2012; 
20.  Diaz \& Branch 1997;  21. Mr\'{o}z P. et al. 2015;  \\22. Schaefer 2021;  23. Schaefer 2009;  24. Wagner et al. 2012
\end{flushleft}

\begin{flushleft}
Evidences:	
A.	The reported periodicity is not coherent, and hence is not orbital;~~
B.	Multiple possible periods and no reliable way to distinguish the orbital period;~~
C.	The data were never published, and the claimed $P$ is given with too little information to inspire confidence;~~
D.	The classification of the system as having a real TNR event is questionable;~~
E.	Counterpart is too faint for any period test with data from any archival source;~~
F.	Claimed periodicity is not significant;~~
G.	Conflicting period claims with comparable evidence mean that neither claim can be considered reliable;~~
H.	Claimed periodicity is not seen in other data sets that should have seen the modulation as claimed;~~
I.	Claimed period seen with nova near maximum (i.e., before the transition), when the nova shell is optically thick, the central binary is completely hidden.;~~
J.	Claimed periodicity is not seen in $TESS$, whereas any such modulation should have been easily seen;~~
K.	The reported semi-regular variability is not any orbital period;~~
L.	The reported periodicity was caused by an artefact introduced into the standard light curve as presented by {\it MAST}.~~
M.	Confusion with identity of nova counterpart.
\end{flushleft}

\end{table}

\section{Listing of All Known Reliable Nova Periods}

The latest list of nova periods is from a year ago, with the good work of Fuentes-Morales et al. (2021), featuring 92 $P$ values.  This list has systematically excluded novae with $P$$>$2 days, all recurrent novae, and has missed a number of reliable $P$ values in the older literature and in the post-submission literature.  And now, this list needs to be updated with my 49 new periods, plus the many improvements and rejections I have found with the modern survey data.  I am in a good position to produce an exhaustive and comprehensive catalog of nova orbital periods.  This section collects all available $P$ measures for novae.  I have a total of 156 reliable nova periods.  

My list does not include `symbiotic novae' (Kenyon 1986; Miko{\l}ajewska 2010), that is the events with very low amplitudes (typically 3 magnitudes and always $<$7 mag), very slow risetimes (months to years), and very long durations (typically years to decades).  These symbiotic novae are sharply distinct from the ordinary thermonuclear-runaway novae that are collected in this paper, in terms of their light curve properties, orbital periods, demographics, dominant physical mechanisms, and evolutionary paths.  

Table 6 presents my final listing of all 156 novae with {\it reliable} $P$ measures.  The first three columns give the GSVC name for the nova (in order of increasing $P$), the year of eruption (``RN" for recurrent novae), and the light curve class from Strope et al. (2010).  The next column gives the orbital period $P$ in days, plus a number in square-brackets pointing to the appropriate reference as tabulated in the footnote.  The last column gives a variety of properties and classes that the nova belongs to, most of them related to the orbital period and its changes.

In the comments column, the 5 novae with $P$$<$0.071 days are notated with ``$<$Gap".  The 5 novae with 0.071$<$$P$$<$0.111 days are inside the Period Gap and notated with ``InGap".  The 28 novae with 0.60$<$$P$$<$10 days must have a subgiant companion star (``SubG").  The 20 novae with $P$$>$10 days must have red giant companions (``RG").  The many systems with apparent eclipses are marked with ``Ecl".  Those novae with coherent non-orbital periods are identified with a ``NonOrbP".  The V1500 Cyg stars (which are greatly brighter long after eruption than pre-eruption, see Schaefer \& Collazzi 2010) are identified by ``V1500".  The novae for which I have measured the century-long $O-C$ curve to get the evolutionary period changes are marked with ``$\dot{P}$".  The novae for which I have measured the change in orbital period across the nova eruption are marked ``$\Delta P$".  The asynchronous polars (``AsyncP") have the spin period and $P$ get out-of-synchronization due to the eruption.  Intermediate polars are marked as ``IP".  Novae for which a real dwarf nova event is seen in quiescence are marked with ``DN".  The exciting daily $10^{39}$ erg superflares on V2487 Oph are marked with ``Superflare".  Neon novae are marked with ``Neon" or ``Ne", while those with a resolved shell of ejecta are labelled ``Shell".  The novae with mystifying pre-eruption rises are labelled ``PreERise".  And one nova, V458 Vul, is spotted inside a newly-formed planetary nebula (``InPNeb").  Some novae have very large amplitude variations in quiescence (``LAmpVar").  The novae detected by {\it Fermi} as $\gamma$-ray sourvces are marked with ``$\gamma$ray" or just ``$\gamma$.  ``Bulge" points to novae that are in our Milky Way's bulge population.  These lists are likely to not be exhaustive, even though most systems in each class are identified.  With this, the one line for each nova gives the reader a quick idea of the characteristics and peculiarities, with each nova's `personality'.

\begin{table}
	\centering
	\caption{All Reliable Orbital Periods for Novae}
	\begin{tabular}{lllll} 
		\hline
		Nova  &  Year  &  LC class  &  $P$~(days)~[Ref.]  &  Comments  \\
		\hline
RW UMi	&	1956	&	...	&	0.05912  [2]	&	$<$Gap, V1500	\\
GQ Mus	&	1983	&	P(45)	&	0.059365  [3]	&	$<$Gap, V1500	\\
CP Pup	&	1942	&	P(8)	&	0.06126454  [4]	&	$<$Gap, V1500, IP?, Shell	\\
IL Nor	&	1893	&	S(108)	&	0.06709  [2]	&	$<$Gap	\\
V458 Vul	&	2007	&	J(20)	&	0.06812255  [5]	&	$<$Gap, InPNeb	\\
T Pyx	&	RN	&	P(62)	&	0.07623361  	&	InGap, Ecl, $\Delta P$, $\dot{P}$,	\\
	&		&		&		&	~~PreERise, V1500	\\
V1974 Cyg	&	1992	&	P(43)	&	0.08125873  [6]	&	InGap, V1500, Neon, Shell	\\
RS Car	&	1895	&	J(80)	&	0.082436  	&	InGap, NonOrbP	\\
DD Cir	&	1999	&	P(16)	&	0.09746  [7]	&	InGap, Ecl, IP?	\\
V Per	&	1887	&	...	&	0.10712347  [8]	&	InGap, Ecl	\\
V597 Pup	&	2009	&	S(6)	&	0.11119  [9]	&	Ecl, IP?	\\
QU Vul	&	1984	&	P(36)	&	0.1117648  [10]	&	Ecl, Shell, Neon	\\
CQ Vel	&	1940	&	S(50)	&	0.11272  [2]	&		\\
V5627 Sgr	&	~1995	&	...	&	0.117161  [11]	&		\\
V2214 Oph	&	1988	&	S(89)	&	0.117515  [12]	&	Neon, Bulge	\\
V630 Sgr	&	1936	&	S(11)	&	0.1179304  [11]	&	Ecl, Neon	\\
V351 Pup	&	1991	&	P(26)	&	0.1182  [13]	&	Shell, Neon	\\
V5116 Sgr	&	2005	&	S(26)	&	0.1237444  [11]	&	Ecl, Bulge	\\
V4633 Sgr	&	1998	&	P(44)	&	0.1255667  [11]	&	V1500, Bulge	\\
V363 Sgr	&	1927	&	S(64)	&	0.126066  [2]	&		\\
DN Gem	&	1912	&	P(35)	&	0.127844  [14]	&		\\
PW Vul	&	1984	&	J(116)	&	0.1285753  	&	NonOrbP, Shell	\\
XX Tau	&	1927	&	D(43)	&	0.1293567  	&	NonOrbP	\\
YZ Ret	&	2020	&	P(22)	&	0.1324539  	&	NonOrbP, Neon, $\gamma$	\\
V4742 Sgr	&	2002	&	S(23)	&	0.1336159  [11]	&	Ecl, Bulge	\\
V1494 Aql	&	1999	&	O(16)	&	0.1346161  [15]	&	Ecl, Neon	\\
V2574 Oph   	&	2004	&	S(41)	&	0.1350862  	&	NonOrbP	\\
V5585 Sgr   	&	2010	&	O(25)	&	0.137526  [11]	&	Ecl, Bulge	\\
V603 Aql	&	1918	&	O(12)	&	0.138201  [16]	&	Shell	\\
V728 Sco	&	1862	&	...	&	0.13833866  [2]	&	Ecl	\\
V1668 Cyg	&	1978	&	S(26)	&	0.1384  [17]	&	Ecl, Neon	\\
DY Pup	&	1902	&	...	&	0.13952  [2]	&	Ecl	\\
V1500 Cyg	&	1975	&	S(4)	&	0.139617  [18]	&	V1500, AsyncP, Shell	\\
	&		&		&		&	~~PreERise, Neon?	\\
RR Cha	&	1953	&	S(60)	&	0.1401  [19]	&	Ecl	\\
V909 Sgr	&	1941	&	...	&	0.14286  [20]	&	Ecl, Neon	\\
RR Pic	&	1925	&	J(122)	&	0.145023762  	&	$\Delta P$, $\dot{P}$, Shell	\\
CP Lac	&	1936	&	S(9)	&	0.145143  [16]	&	Shell	\\
V2468 Cyg	&	2008	&	S(20)	&	0.14525  [21]	&		\\
V500 Aql	&	1943	&	S(43)	&	0.145259  	&		\\
V382 Vel	&	1999	&	S(13)	&	0.1461  	&	Neon	\\
NQ Vul	&	1976	&	D(50)	&	0.1462568  	&	Shell	\\
V533 Her	&	1963	&	S(43)	&	0.147  [22]	&	IP?, Shell, PreERise	\\
FM Cir	&	2018	&	J(85)	&	0.1497672  	&		\\
V5113 Sgr	&	2003	&	J(48)	&	0.150015  [11]	&	Bulge	\\
V999 Sgr	&	1910	&	J(160)	&	0.1518412  [11]	&	Bulge	\\
V1674 Her	&	2021	&	S(2)	&	0.15302  [23]	&	IP, $\gamma$ray	\\
V4579 Sgr	&	1986	&	...	&	0.153561  [11]	&	Ecl	\\
V992 Sco	&	1992	&	D(120)	&	0.1536  [7]	&		\\
WY Sge	&	1783	&	...	&	0.1536345  [2]	&	Ecl, DN?	\\
V2467 Cyg	&	2007	&	O(20)	&	0.153789  	&		\\
X Cir	&	1926	&	...	&	0.15445953  [2]	&	Ecl	\\
V357 Mus	&	2018	&	D(32)	&	0.155163  	&		\\
OY Ara	&	1910	&	S(80)	&	0.15539  [2]	&	Ecl	\\
V1493 Aql	&	1999	&	C(50)	&	0.156  [24]	&		\\
V1369 Cen	&	2013	&	D(65)	&	0.156556  	&	$\gamma$ray	\\
V5582 Sgr	&	2009	&	J(90)	&	0.156604  [11]	&	Bulge	\\
V2572 Sgr	&	1969	&	S(44)	&	0.157038  	&		\\
V598 Pup	&	2007	&	...	&	0.162874  	&	NonOrbP	\\
V339 Del	&	2013	&	PP(29)	&	0.162941  	&	$\gamma$ray	\\
DO Aql	&	1925	&	F(900)	&	0.167762  [25]	&	Ecl	\\
V390 Nor	&	2007	&	...	&	0.171326  	&		\\
V849 Oph	&	1919	&	F(270)	&	0.17275611  [2]	&	Ecl, Neon	\\

		\hline
	\end{tabular}
\end{table}

\begin{table}
	\centering
	\contcaption{All Reliable Orbital Periods for Novae}
	\label{tab:continued}
	\begin{tabular}{lllll}
		\hline
		Nova  &  Year  &  LC class  &  $P$~(days)~[Ref.]  &  Comments  \\
		\hline

HS Pup	&	1963	&	S(65)	&	0.178641  	&		\\
V5558 Sgr	&	2007	&	J(157)	&	0.185808  	&	Bulge	\\
V1405 Cas	&	2021	&	J(175)	&	0.1883907  	&	$\gamma$ray	\\
V825 Sco	&	1964	&	...	&	0.191659  [11]	&	Ecl, Neon, Bulge	\\
DQ Her	&	1934	&	D(100)	&	0.1936208997  	&	Ecl, $\Delta P$, $\dot{P}$,	\\
	&		&		&		&	~~IP, Shell	\\
CT Ser	&	1948	&	...	&	0.195  [26]	&		\\
AT Cnc	&	c.1686	&	...	&	0.201634  [27]	&	DN	\\
V1186 Sco	&	2004	&	J(62)	&	0.202968  	&	Ecl	\\
T Aur	&	1891	&	D(84)	&	0.2043783  [28]	&	Ecl, Shell	\\
V446 Her	&	1960	&	S(42)	&	0.207  [22]	&	DN, Shell	\\
IM Nor      	&	RN	&	P(80)	&	0.2071655  [30]	&	Ecl, $\dot{P}$	\\
V4745 Sgr	&	2003	&	J(190)	&	0.20782  [31]	&	IP?	\\
HZ Pup	&	1963	&	J(70)	&	0.212  [32]	&	IP	\\
V1213 Cen	&	2009	&	C(30)	&	0.21201  [33]	&	DN	\\
HR Del	&	1967	&	J(231)	&	0.21416215  	&	$\Delta P$, $\dot{P}$	\\
	&		&		&		&	~~Shell, Neon	\\
AR Cir	&	1906	&	J(330)	&	0.2143  [34]	&		\\
V5588 Sgr   	&	2011	&	O(77)	&	0.21432  [11]	&		\\
NR TrA	&	2008	&	J(61)	&	0.2192  [35]	&	Ecl	\\
CN Vel	&	1905	&	S(850)	&	0.2202  [34]	&		\\
V365 Car	&	1948	&	S(530)	&	0.22369  [2]	&		\\
V705 Cas	&	1993	&	D(67)	&	0.228284  	&	Shell	\\
V1039 Cen	&	2001	&	J(174)	&	0.247  [36]	&	IP?	\\
V1425 Aql	&	1995	&	S(79)	&	0.2558  [37]	&	IP?	\\
V2615 Oph   	&	2007	&	S(48)	&	0.272339  [11]	&		\\
V972 Oph	&	1957	&	J(176)	&	0.279641  	&		\\
V4743 Sgr	&	2002	&	S(17)	&	0.2799  [38]	&	IP, Neon	\\
BY Cir	&	1995	&	P(124)	&	0.2816  [7]	&	Ecl	\\
V2540 Oph	&	2002	&	J(115)	&	0.284781  [39]	&	Ecl	\\
V1059 Sgr	&	1898	&	...	&	0.2861  [40]	&		\\
Z Cam	&	<c.700	&	...	&	0.289841  [41]	&	DN	\\
V959 Mon	&	2012	&	...	&	0.29585  [42]	&	Neon, $\gamma$ray	\\
V838 Her	&	1991	&	P(4)	&	0.297635  [43]	&	Ecl, Neon	\\
V1174 Sgr	&	1952	&	...	&	0.3090452  [11]	&	Bulge	\\
V2275 Cyg	&	2001	&	S(8)	&	0.31449  [44]	&		\\
BT Mon	&	1939	&	F(182)	&	0.3338149  	&	Ecl, $\Delta P$, $\dot{P}$,	\\
	&		&		&		&	~~Shell	\\
V2677 Oph   	&	2012	&	S(11)	&	0.344295  [11]	&		\\
QZ Aur	&	1964	&	S(25)	&	0.35749703  	&	Ecl, $\Delta P$, $\dot{P}$	\\
V1375 Cen	&	2008	&	...	&	0.3604  [45]	&		\\
V549 Vel	&	2017	&	J(118)	&	0.4031692  	&	$\gamma$ray	\\
Q Cyg	&	1876	&	S(11)	&	0.42036  [46]	&		\\
V2860 Ori	&	2019	&	P(13)	&	0.422579  	&	Plateau(2021)	\\
V356 Aql	&	1936	&	J(140)	&	0.4265059  	&		\\
V719 Sco	&	1950	&	D(24)	&	0.43639  	&	Bulge	\\
GI Mon	&	1918	&	S(23)	&	0.4470645  	&	Ecl, DN, IP?	\\
V1330 Cyg	&	1970	&	S(217)	&	0.50	&		\\
J17014 4306	&	c.1437	&	...	&	0.5340055  [47]	&	Ecl, DN	\\
V841 Oph	&	1848	&	S(140)	&	0.601304  [46]	&	SubG	\\
CI Aql	&	RN	&	P(32)	&	0.61836092  	&	SubG, Ecl, 	\\
	&		&		&		&	~~$\Delta P$, $\dot{P}$	\\
V368 Aql	&	1936	&	S(42)	&	0.6905093  [48]	&	SubG, Ecl	\\
V723 Cas	&	1995	&	J(299)	&	0.693265  [49]	&	SubG, V1500,	\\
	&		&		&		&	~~Neon, Shell	\\
V373 Sct	&	1975	&	J(79)	&	0.819099  	&	SubG	\\
V726 Sgr	&	1936	&	S(95)	&	0.822812  [11]	&	SubG, Neon, Bulge	\\
V400 Per	&	1974	&	...	&	0.826387  	&	SubG	\\
QY Mus	&	2008	&	D(100)	&	0.901135  	&	SubG, NonOrbP	\\
HR Lyr	&	1919	&	...	&	0.905778  	&	SubG	\\
V3732 Oph	&	2021	&	...	&	0.940043  [50]	&	SubG, Bulge	\\
CP Cru	&	1996	&	S(10)	&	0.944  [7]	&	SubG, Ecl, Neon	\\
U Sco	&	RN	&	PP(3)	&	1.23055183  	&	SubG, Ecl, Neon?	\\
	&		&		&		&	~~$\Delta P$, $\dot{P}$	\\

		\hline
	\end{tabular}
\end{table}

\begin{table}
	\centering
	\contcaption{All Reliable Orbital Periods for Novae}
	\label{tab:continued}
	\begin{tabular}{lllll}
		\hline
		Nova  &  Year  &  LC class  &  $P$~(days)~[Ref.]  &  Comments  \\
		\hline

V2487 Oph	&	RN	&	P(8)	&	1.24  	&	SubG, Superflare,	\\
	&		&		&		&	~~Bulge	\\
V697 Sco	&	1941	&	...	&	1.26716  	&	SubG, Ecl, Bulge	\\
V2674 Oph   	&	2010	&	S(31)	&	1.30207  [11]	&	SubG, Ecl, Bulge	\\
V2109 Oph	&	1969	&	...	&	1.32379  	&	SubG, DN, Bulge	\\
X Ser	&	1903	&	S(730)	&	1.478  [22]	&	SubG, DN	\\
V394 CrA	&	RN	&	P(5)	&	1.515682  	&	SubG, Ecl, $\dot{P}$,	\\
	&		&		&		&	~~LAmpVar, Bulge	\\
V5589 Sgr	&	2012	&	S(13)	&	1.5923  [11]	&	SubG, Ecl, Bulge	\\
HV Cet	&	2008	&	...	&	1.7718  [51]	&	Neon, SubG, PreERise	\\
V1370 Aql	&	1982	&	D(29)	&	1.9581  	&	Neon, SubG	\\
GK Per	&	1901	&	O(13)	&	1.996803  [52]	&	SubG, DN, IP,	\\
	&		&		&		&	~~Shell, Neon	\\
KT Eri	&	2009	&	PP(14)	&	2.61595  	&	SubG, LAmpVar	\\
V392 Per	&	2018	&	P(11)	&	3.21997  	&	SubG, DN, Neon, $\gamma$	\\
V407 Lup	&	2016	&	S(8)	&	3.62  	&	SubG, NonOrbP, $\gamma$	\\
FS Sct	&	1952	&	...	&	5.7  	&	SubG	\\
V1017 Sgr	&	1919	&	S(130)	&	5.78629  	&	SubG, DN, $\Delta P$, $\dot{P}$	\\
V723 Sco	&	1952	&	S(23)	&	7.2	&	SubG, Bulge	\\
V1016 Sgr	&	1899	&	...	&	13.4  	&	RG, NonOrbP, IP?	\\
V794 Oph	&	1939	&	J(220)	&	18  	&	RG, Bulge	\\
V977 Sco	&	1989	&	...	&	26.2  	&	RG, Neon, Bulge	\\
V4338 Sgr	&	1990	&	...	&	29.481916  	&	RG, Bulge	\\
GR Sgr	&	1924	&	...	&	29.4956  	&	RG	\\
V5580 Sgr	&	2008	&	...	&	34  	&	RG, Bulge	\\
V3645 Sgr	&	1970	&	...	&	39.5  	&	RG, Bulge?	\\
V1313 Sco	&	2011	&	S(18)	&	50  	&	RG, Bulge	\\
V1535 Sco	&	2015	&	S(20)	&	50  	&	RG, Bulge	\\
EU Sct	&	1949	&	S(42)	&	68  	&	RG	\\
T CrB	&	RN	&	S(6)	&	227.532  [53]	&	RG, PreERise,	\\
	&		&		&		&	~~$\Delta P$, $\dot{P}$	\\
V1172 Sgr	&	1951	&	...	&	255  	&	RG, Bulge	\\
KY Sgr	&	1926	&	S(109)	&	294  	&	RG, Bulge	\\
RS Oph	&	RN	&	P(14)	&	453.6  [54]	&	RG, $\gamma$, PostDip	\\
V1534 Sco	&	2014	&	S(9)	&	520  	&	RG, Bulge	\\
V1657 Sco	&	2017	&	...	&	650  	&	RG, Bulge?	\\
V3890 Sgr	&	RN	&	S(14)	&	747.6  [55]	&	RG, $\gamma$ray, Bulge	\\
V5581 Sgr	&	2009	&	...	&	1660  	&	RG, Bulge	\\
V3664 Oph	&	2018	&	...	&	2250  	&	RG, Bulge	\\
V745 Sco	&	RN	&	P(9)	&	2440  	&	RG, $\gamma$ray, Bulge	\\

		\hline
	\end{tabular}

\begin{flushleft}
References:  If no reference is given, then the source of the quoted $P$ is this paper, with details and references given in Tables 1--5 and in the text.  2. Fuentes-Morales et al. 2021;  3. Narloch et al. 2014;  4. Bianchini et al. 2012;  5. Rodr\'{i}guez-Gil et al. 2010; 
6. Olech 2002;  7. Woudt \& Warner 2003;  8. Shafter \& Misselt 2006;  9. Warner \& Woudt 2009;  10. Shafter et al. 1995;  11. Mr\'{o}z P. et al., 2015;  
12. Baptista et al. 1993;  13. Woudt \& Warner 2001;  14. Retter et al. 1999;  15. Kato et al. 2004;  16. Peters \& Thorstensen 2006;  17.Kaluzny 1990;  18. Pavlenko et al. 2018
19. Woudt \& Warner 2002;  20. Tappert et al. 2013;  21. Chochol et al. 2013;  22. Thorstensen \& Taylor 2000;  23. Patterson et al. 2021;  
24. Dobrotka et al. 2006a;  25. Shafter et al. 1993;  26. Ringwald et al. 2005; 27. Shara et al. 2017;  28. Dai \& Qian 2010;  29. Thorstensen \& Taylor 2000;  
30. Patterson et al. 2022;  31. Dobrotka et al. 2006b;  32. Thorstensen et al. 2017;  33. Mr\'{o}z P., et al. 2016;  34. Tappert et al. 2013;  35. Walter 2015;  
36. Woudt et al. 2005;  37. Retter et al. 1998;   38. Kang et al. 2006;  39. Ak et al. 2005;  40. Thorstensen et al. 2010;  41. Shara et al. 2012;  
42. Munari et al. 2013;  43. Ingram et al. 1992;  44. Balman et al. 2005;  45. Patterson et al. 2010;  46. Peters \& Thorstensen 2006;  47. Shara et al. 2017;  
48. Marin \& Shafter 2009;  49. Ochner et al. 2015;  50. Mr\'{o}z P. et al. 2021;  51. Beardmore et al. 2012;  52. Morales-Rueda et al. 2002;  
53. Leibowitz et al. 1997;  54. Brandi et al. 2009;  55. Miko{\l}ajewska et al. 2021
\end{flushleft}

\end{table}

\section{Period Histograms}

The premier application of my exhaustive compilation of all known reliable nova $P$ values is to construct a histogram of the period distribution.  Such histograms have been of historical importance for establishing and quantifying the Period Gap.  Important analyses of the histogram have been to measure the shortest $P$ ($P_{\rm min}$ or $P_{\rm bounce}$) and the boundaries of the Period Gap (from $P_{\rm gap, -}$ to $P_{\rm gap, +}$), for detailed comparison with CV-evolution models.  A typical and critical result is to demonstrate that there must be some residual magnetic braking for novae below the Period Gap (Knigge et al. 2006).  The schematic nova histogram, used by everyone, has few novae below a 2--3 hour Period Gap, a high peak from roughly 3--6 hours, and no one pays any attention to $P$$>$1 day systems.  Further, below the Period Gap, the simple calculation (with only General Relativity providing the loss of angular momentum in the binary) gives $P_{\rm min}$ of 65 minutes (0.045 days).  This schematic picture is only approximately correct.  Knigge et al. (2006) has made a thorough analysis of the period histogram for all CVs (novae, plus nova-likes, plus dwarf novae) to get $P_{\rm min}$=76.2$\pm$1.0 minutes, $P_{\rm gap,-}$=2.15$\pm$0.03 hours, and $P_{\rm gap,+}$=3.18$\pm$0.04 hours.  Various measures of the critical periods are tabulated in Table 7.

\begin{table}
	\centering
	\caption{Critical periods from the histograms}
	\begin{tabular}{llll} 
		\hline
		CVs  &  $P_{\rm min}$ (days)  &  $P_{\rm gap,-}$ (days)  & $P_{\rm gap,+}$ (days)  \\
		\hline
Schematic	&	0.045	&	0.083	&	0.125	\\
All CVs	&	0.0529	&	0.0896	&	0.1325	\\
Novae	&	0.059	&	0.071	&	0.111	\\
Nova-likes	&	0.054	&	0.081	&	0.131	\\
Dwarf Novae	&	0.0525	&	0.0925	&	0.141	\\
		\hline
	\end{tabular}
\end{table}

Nova period histograms have progressed from 10 novae (Patterson 1984) in the early days to 79 novae with correct periods (Fuentes-Morales et al. 2021) from just a few months ago.  Now, the number of reliable nova $P$ in my listing (156) is double the previous listing.

\begin{figure}
	\includegraphics[width=1.01\columnwidth]{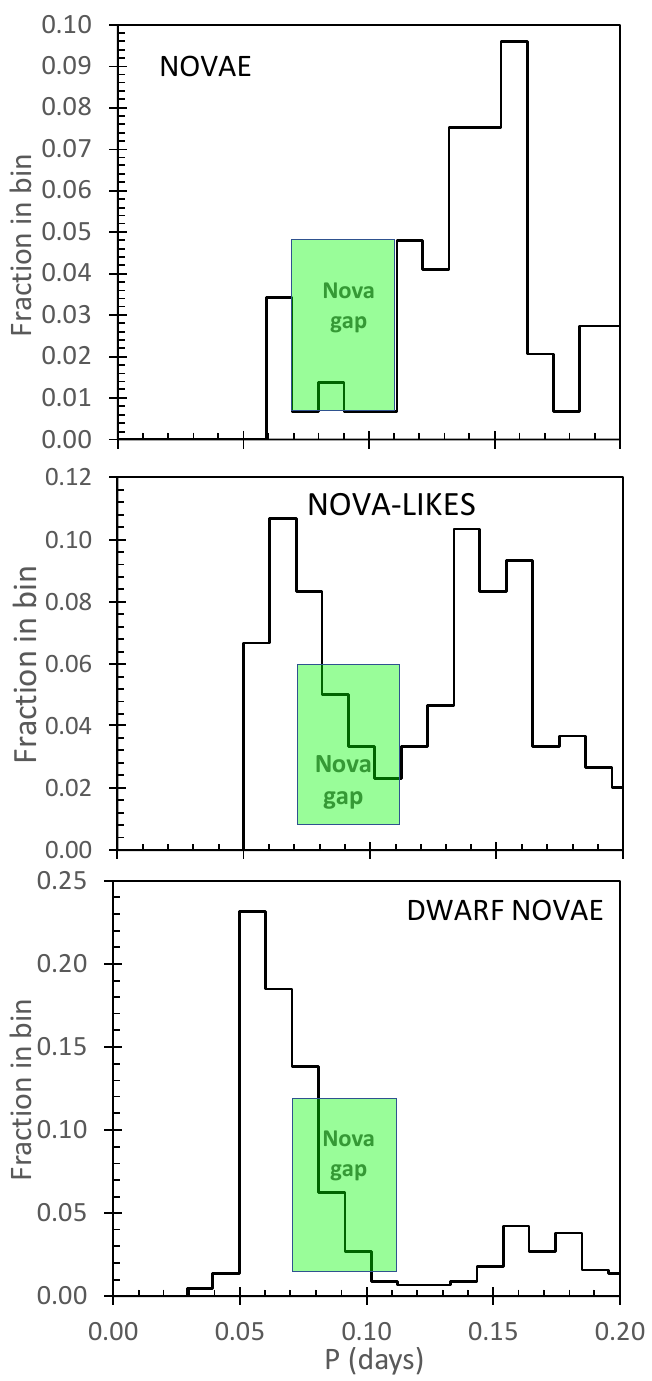}
    \caption{The period histogram for 156 novae, 300 novalikes, and 449 dwarf novae.  This linear plot for $P$$<$0.2 days is made to emphasize the Period Gap and the minimum period.  The Period Gap for novae is 0.071--0.111 days (1.70--2.66 hours), as illustrated by the horizontal extent of the light-green box.  The minimum $P$ for novae is close to 0.059 days (85.0 minutes).  The Period Gap for novalikes is from 0.081--0.131 days (1.94--3.14 hours).  The Period Gap for dwarf novae is from 0.0925--0.141 days (2.22--3.38 hours).  Importantly and surprisingly, the gap for the novae is greatly different from the gaps for dwarf novae and the nova-like systems.  This is surprising, as there has been no realization, expectation, or prediction that the position of the gap will change with the CV class.}
\end{figure}

Fig. 4 shows the period histogram for the novae, with this being a close-up for $P$ less than 0.20 days, all on a linear scale.  The gap covers 0.071--0.111 days, or 1.70--2.66 hours.  Further, the lower limit on $P$ for novae appears to be close to 0.059 days.

\begin{figure}
	\includegraphics[width=1.01\columnwidth]{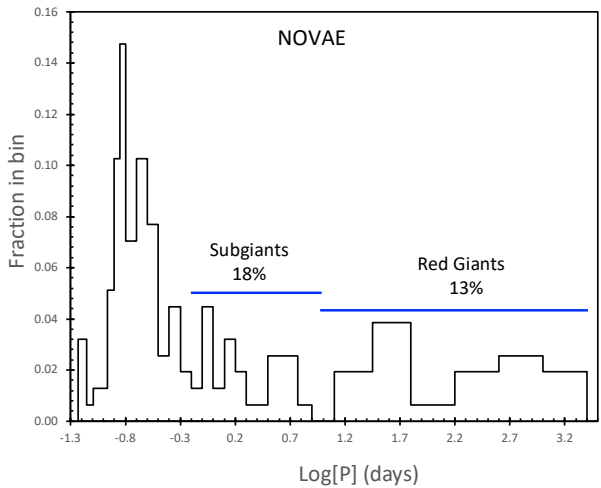}
    \caption{The period histogram for novae on a logarithmic scale.  The scale covers all the nova periods, from 0.059--2440 days.  Novae with a subgiant companion star will have periods from approximately 0.6--10 days, while those with a red giant companion will have periods of $>$10 days or so.  The 28 novae with subgiant companions comprise 18 per cent, while the 20 novae with red giant companions comprise 13 per cent.}
\end{figure}

Novae have a very broad range of orbital periods (0.059 to near 2440 days), and a linear plot will be misleading for displaying long period systems.  A solution is to plot the histogram for logarithmic bins, where the vertical axis gives the count in each bin, as in Fig. 5.  The distribution above the gap has a high peak near 0.15 days, then an exponential-like decline to longer periods.  The novae with subgiant companions comprises 18 per cent of all novae, while the systems with red giant companions comprise 13 per cent of all novae.  The reason for calling these out is to express the always-ignored fact that a large fraction of nova systems have evolved companions.  With 31 per cent of all novae having evolved companions, they must be accounted for in all models of CV evolution.  I know of no theorist, modeler, or observer that recognized, considered, or modeled the large fraction of evolved companions or the evolution of these systems\footnote{A cause and symptom of this neglect is that models of CV evolution are not applicable, with Magnetic Braking only considering systems with $P$$<$7 hours (Rappaport, Verbunt, \& Joss 1985; Knigge et al. 2011), the calculations for the Hibernation model only considering systems with $P$$<$6 hours (Shara et al. 2018), and the Consequential Angular Momentum Loss (CAML) model only considering CVs with $P$$<$10 hours (Schreiber, Zorotovic \& Wijnen 2016).  Further, a cause and symptom of ignoring the novae with evolved companions is that the prior compilations have cutoffs for novae with $P$$>$2 days (e.g., Patterson 1984; Fuentes-Morales et al. 2021) or $P$$>$5.7 days (Ritter \& Kolb 2003).}.

Nova-like CVs are systems that appear to be ordinary nova systems for which no nova eruption has been seen.  The best and largest compilation of nova-like CV periods is the Ritter \& Kolb catalog.  I have taken all 300 nova-like CVs with reliable periods.  The histogram is plotted in the middle panel of Fig. 4.  Tall peaks appear on either side of a well defined Period Gap.  The numbers have a sharp drop from a high value down to zero at a period of 0.054 days, which I identify as $P_{\rm min}$.   The gap for nova-like CVs is from 0.081 to 0.131 days, or 1.94--3.14 hours.  Critically, the upper edge of this gap is greatly and significantly different from the upper edge of the nova gap.  This can be seen emphatically in Fig. 4, where the light-green box labelled `Nova gap' shows the period range of the Period Gap for novae.  There is only modest overlap with the novae.

Dwarf novae (DNe) systems are largely identical to nova systems.  The Ritter \& Kolb catalog has 449 dwarf novae with reliable periods.  The histogram is plotted in the lower panel of Fig. 4.  This is dominated by a tall peak below the Period Gap, with a well-defined gap.   The gap for dwarf novae is from 0.0925 to 0.141 days, or 2.22--3.38 hours.  The upper edge of this gap is greatly and significantly different from the nova gap.  This can be seen emphatically in Fig. 4, where there is only modest overlap with the novae.  The histogram shows a sharp drop at 0.0525 days, and I identify this as $P_{\rm min}$.  However, 9 DNe have periods 0.038--0.051 days.

The minimum period is just a function of the angular momentum loss, which is schematically from General Relativity alone, and should be the same for novae, DNe, and nova-likes.  Further, the physical mechanisms of all the CVs really should be identical.  $P_{\rm min}$ should be identical across all CVs.  But looking at the histograms, novae have zero systems with $P$$<$0.059 days, while both nova-like CVs and DNe have many systems with substantially shorter periods.  It looks like novae have a different $P_{\rm min}$ than other CVs.  However, detailed statistical analyses (including a Kolmogorov-Smirnov Test) of the three CV classes show that the systems below the Period Gap do not have different parent populations at more than the 3-$\sigma$ confidence level.

The lower edge of the Period Gap appears similar for novae, nova-like CVs, and DNe.  The histograms all show a fall-off from below the gap down into the middle of the gap.  The lack of a sharp drop-off in the histograms makes for $P_{\rm gap,-}$ to be poorly defined.  A Kolmogorov-Smirnov Test for the period distributions from 0.061--0.110 days (i.e., definitely above $P_{\rm min}$ and below $P_{\rm gap,+}$ for all the types of CVs) shows that the novae, DNe, and nova-likes are consistent with being from the same parent population.

The upper edge of the Period Gap is sharply defined in all the $P$ histograms.  The $P_{\rm gap,+}$ values and their formal one-sigma error bars are 0.111$\pm$0.007 days for novae, 0.131$\pm$0.002 days for the nova-likes, and 0.141$\pm$0.003 days for the dwarf novae.  Detailed statistical analysis (including the Kolmogorov-Smirnov test) of the three $P$ distributions from 0.095--0.18 days can determine whether the distributions come from the same parent population.  The probability is 0.00022 that the novae and nova-likes are from the same parent population, 0.00026 for the novae and dwarf novae, and 0.00039 for the nova-likes and the dwarf novae.  So the three CV classes have significantly different $P_{\rm gap,+}$.  But formal statistics are not needed, as the $P_{\rm gap,+}$ values are greatly different for each of the three CV classes, with this being easily seen in Fig. 4.

\section{New and Critical Open Questions}

{\bf A.}~~How can the novae, nova-likes, and DNe all have significantly different $P_{\rm gap,+}$?  All three CV classes are the same systems (e.g., nova-likes and DNe both have nova eruptions, while old novae are nova-like systems and often have DN events), while at any one $P$ their physical mechanisms must be the same.  The average properties of the groups, such as accretion rate, white dwarf masses, and evolutionary age, vary between the CV classes, and these must be the root of the change in $P_{\rm gap,+}$.  Knigge et al. (2011) explain how $P_{\rm gap,+}$ changes as a function only of the strength of the magnetic braking.  They quantify this as some factor times the magnetic braking as given by a standard parametrization.  For novae, with the upper edge of the gap at 0.111 days, the effective magnetic braking is 0.20$\times$ that of their standard parametrization.  Nova-likes, with $P_{\rm gap,+}$=0.131 days, have a factor of 0.57$\times$, while DNe, with the upper edge of the gap at 0.141 days, have a factor of 0.82$\times$.  Within the standard Magnetic Braking model, the reason why novae have a substantially smaller $P_{\rm gap,+}$ than do DNe, is because their magnetic braking efficiency is near 4$\times$ smaller.  There is a real possibility that the standard Magnetic Braking model might require a significant correction or addition so as to account for the differing $P_{\rm gap,+}$ values.  For example, it is plausible that moderate or large magnetic fields on the white dwarf will affect the gap, and perhaps these effects are correlated with the bulk properties of the three CV classes.

{\bf B.}~~A surprising open question is to understand the existence of the coherent non-orbital periods (see Table 4), with these novae each having 1--3 such periods, with most being perplexing and mysterious in origin.  Five of the novae (V842 Cen, QY Mus, V598 Pup, XX Tau, and PW Vul) have coherent, long-lasting, and highly significant periodicities from 0.12--4.1 days, which are certainly {\it not} the orbital period.  Two of the systems (V2574 Oph and V407 Lup) have fast periodicities that are stable and coherent, with the only plausible good `clock' being the white dwarf spin, yet for which they are not the spin period, and there must be some unknown clocking mechanism.  The unique new phenomena for YZ Ret can be speculated to be associated with some exotic superhump condition, with no precedent or prior theory.  These modulations are mysterious because their coherence forces them to each be tied to some accurate `clock' in the system, but they cannot be either orbital or spin, while no other highly-coherent clocks can be invoked.  Throughout astrophysics, whenever a periodicity is discovered, {\it physics} can be applied to the system, while I would be hopeful that an understanding of these coherent non-orbital periods can provide the physics of some previously unknown mechanisms operating in nova binaries.  To the best of my knowledge of the literature, no one has ever recognized or highlighted this basic problem.  

{\bf C.}~~Novae with evolved companions are now seen to make up 31 per cent of all Milky Way novae.  These novae have been completely ignored by modelers.  No idea has been published as to the past and future changes in the accretion rate, $\dot{P}$, and stellar masses.  Is the accretion being driven by a decrease in $P$, or by the evolutionary expansion of the companion?  Indeed, the dominant mechanism for angular momentum loss in the binary is not known.  A conundrum recognized in this paper is that the novae with red giant companions are 75 per cent in the bulge population, with this being difficult to understand in detail, especially when the novae with red giant companions in the Andromeda Galaxy comprise $\sim$30 per cent of the observed novae and these all appear to be in the {\it disc} population (Williams et al. 2016).  Another unaddressed question is whether the currently observed systems have only recently come into contact.  Why are these novae so sharply different from symbiotic novae?  With some evolved companions likely being more massive than the white dwarf, will the simple mass transfer drive runaway accretion?  Is the white dwarf gaining or losing mass over each eruption cycle, with this leading into the question of whether CVs are the solution to the highly-important Type Ia supernova Progenitor Problem?

\section{Acknowledgements}

I thank Rebekah Hounsell for helping me to get working the {\sc Lightkurve} program available from MAST.  The AAVSO provides the unique tools that formed the backbone of my work, including variable star cataloguing (VSX), their huge International Database of light curves, their finder chart facility (VSP), their calibration of comparison stars (APASS), their DFT routine (VSTAR), plus their ability to respond to get new special-purpose data.  My use of the high level data products of the many sky surveys does not well highlight the huge efforts over many years by large teams of workers from $TESS$, ZTF, K2, MAST, AAVSO, SMARTS, ASAS, and OGLE.  Funding for the $TESS$ mission is provided by NASA's Science Mission directorate.  This research made use of {\sc Lightkurve}, a Python package for $Kepler$ and $TESS$ data analysis (Lightkurve Collaboration, 2018).

\section{Data Availability}

All of the photometry data used in this paper are publicly available from the references and links in Section 2.1.


{}

%%%%%%%%%%%%%%%%%%%%%%%%%%%%%%%%%%%%%%%%%%%%%%%%%%
% Don't change these lines
\bsp	% typesetting comment
\label{lastpage}
\end{document}